\documentclass[pra,twocolumn,aps,amssymb,footinbib, 10pt]{revtex4}%

\usepackage{amsfonts}
\usepackage{amsmath}
\usepackage{amssymb}
\usepackage{graphicx}
\usepackage{bbm}
\usepackage{calc}
\usepackage{capt-of}
\usepackage{amsfonts}
\usepackage{amsmath,enumerate}
\usepackage{amssymb}
\usepackage{graphicx}
\usepackage{color} 
\usepackage{amsmath}%
\providecommand{\U}[1]{\protect\rule{.1in}{.1in}}

\newenvironment{proof}[1][Proof]{\noindent\textbf{#1.} }{\ \rule{0.5em}{0.5em}}

\newcommand{\tmop}[1]{\ensuremath{\operatorname{#1}}}

\newcommand{\OK}{\checkmark}

\newcommand{\trace}{\tmop{Tr}}

\renewcommand{\Pr}{\tmop{\bf Pr}}

\newcommand{\comentout}[1]{}

\begin{document}

\title{Unforgeable Noise-Tolerant Quantum Tokens}
\author{F. Pastawski$^{1*}$, N. Y. Yao$^{2}$, L. Jiang$^{3}$, M. D. Lukin$^{2}$, J. I. Cirac$^{1}$}

\affiliation{$^{1}$Max-Planck-Institut fur Quantenoptik, Hans-Kopfermann-Strase 1, Garching, D-85748, Germany}
\affiliation{$^{2}$Physics Department, Harvard University, Cambridge, MA 02138}
\affiliation{$^{3}$Institute for Quantum Information, California Institute of Technology, Pasadena, CA 91125}
\email{fernando.pastawski@mpq.mpg.de}

\date{\today}

\begin{abstract}
\end{abstract}
\maketitle

\preprint{APS/123-QED}

{\bf The realization of devices which harness the laws of quantum mechanics represents an exciting challenge at the interface of modern technology and fundamental science\cite{ChuangBook, gisin_quantum_2002}. An exemplary paragon of the power of such quantum primitives is the concept of ``quantum money'' \cite{wiesner_conjugate_1983}. A dishonest holder of a quantum bank-note will invariably fail in any forging attempts; indeed, under assumptions of ideal measurements and decoherence-free memories such security is guaranteed by  the no-cloning theorem \cite{wootters_single_1982}. In any practical situation, however, noise, decoherence and operational imperfections abound. Thus, the development of secure ``quantum money''-type primitives  capable of tolerating realistic infidelities is of both practical and fundamental importance. Here, we propose a novel class of such protocols and demonstrate their tolerance to noise; moreover, we prove their rigorous security by determining tight fidelity thresholds. Our proposed protocols require only the ability to prepare, store and measure single qubit quantum memories, making their experimental realization accessible with current technologies \cite{Hume07, Wendin03, Maurer11}. }

Recent extensions to Wiesner's original ``quantum money'' protocol \cite{wiesner_conjugate_1983} have garnered significant  interest  \cite{aaronson_quantum_2009, lutomirski_breaking_2009, farhi_quantum_2010, lutomirski_component_2011}. One particular extension enables the authentication of quantum tokens via classical public communication with a trusted verifier \cite{gavinsky_quantum_2011}. However, to tolerate noise, the verification process must condone a certain finite fraction of qubit failures; naturally, such a relaxation of the verification process enhances the ability for a dishonest user to forge quantum tokens. It is exactly this interplay which we, here, seek to address, by focusing  on a class of "quantum token"-protocols which involve either direct physical  or classical communication verification of qubit memories.

Our approach to quantum tokens extends the original quantum money primitive\cite{wiesner_conjugate_1983} by ensuring tolerance to finite errors associated with encoding, storage and decoding of individual qubits. 
We denote the tokens within our first primitive as quantum tickets (qtickets);
each qticket is issued by the mint and consists of a unique serial number and $N$ component quantum states, $\rho=\bigotimes_{i}\rho_i$, where each $\rho_i$ is drawn uniformly at random from the set, $\tilde{Q} = \{ \vert + \rangle, \vert - \rangle, \vert +i \rangle, \vert -i \rangle, \vert 0 \rangle, \vert 1 \rangle\}$, of polarization eigenstates of the Pauli spin operators.  The mint secretly stores a classical description of $\rho$, distributed only among trusted verifiers.
In order to redeem a qticket, the holder physically deposits it with a trusted verifier, who
 measures the qubits in the relevant basis. This verifier then requires a minimum fraction, $F_{\tmop{tol}}$, of correct outcomes in order to authenticate the qticket; following validation, the only information returned by the verifier is whether the qticket has been accepted or rejected. 
 
The soundness of a qticket, e.g. the probability that an honest user is successfully verified, depends crucially on the experimental fidelities associated with single qubit encoding, storage and decoding. Thus, for a given qubit $\rho_i$, we define the map, $M_i$, which characterizes the overall fidelity, beginning with the mint's encoding and ending with the verifier's validation; the average channel fidelity\cite{nielsen_simple_2002} is then given by, $F_i = 1/|\tilde{Q}| \sum_{\rho_i} \trace[\rho_i M_i(\rho_i)]$. With this definition, the verification probability of an honest user is,
\begin{equation}\label{eq:NoHonestFail}
p_{\tmop{h}}=\frac{1}{|Q|}\sum_{\rho \in Q} \trace[P_{\tmop{acc}} M(\rho)] \geq 1 - e^{-N D(F_{\tmop{exp}} \| F_{\tmop{tol}})},
\end{equation}
where $Q = \tilde{Q}^{\otimes N}$, $P_{\tmop{acc}}$ represents the projector onto the subspace of valid qtickets, $M = \bigotimes_{i} M_i$, $F_{\tmop{exp}} = 1/N \sum_i F_i$ is the average experimental fidelity, and $D$, the relative entropy, characterizes the distinguishability of two distributions (see Methods for details).  Crucially, so long as the average experimental fidelity associated with single qubit processes is greater than the tolerance fidelity, an honest user is exponentially likely to be verified. 

To determine a tight security threshold, we consider the counterfeiting of a single qticket. 
For a given tolerance fidelity ($F_{\tmop{tol}}$) set by the verifiers, a qticket is only accepted if at least $F_{\tmop{tol}}N$ qubits are validated. 
In the event that a dishonest user attempts to generate two qtickets from a single valid original, \emph{each} must contain a minimum of $F_{\tmop{tol}}N$ valid qubits to be authenticated. 
As depicted in Fig. 1a., in order for each counterfeit qticket to contain $F_{\tmop{tol}}N$ valid qubits, a \emph{minimum} of $(2F_{\tmop{tol}}-1) N$ qubits must have been perfectly cloned. Thus, for a set tolerance fidelity, in order for a dishonest user to succeed, he or she must be able to emulate a qubit cloning fidelity of at least $2F_{\tmop{tol}}-1$. 
Crucially, so long as this fidelity is above that achievable for optimal qubit cloning (2/3) \cite{werner_optimal_1998}, a dishonest user is exponentially unlikely to succeed,
\begin{equation}\label{eq:NoCounterfeiting}
p_{\tmop{d}}=\frac{1}{|Q|}\sum_{\rho \in Q} \trace\left[P_{\tmop{acc}}^{\otimes 2} T(\rho)\right] \leq  e^{-N D( 2F_{\tmop{tol}}-1 \| 2/3)},
\end{equation}
where $T$ represents any completely positive trace preserving qticket counterfeiting map.  
To ensure $2F_{\tmop{tol}}-1 > 2/3$, the tolerance fidelity must be greater than $5/6$, which is precisely the average fidelity of copies produced by an optimal qubit cloning map \cite{werner_optimal_1998}. 
In certain cases, an adversary may be able to sequentially engage in multiple verification rounds; however, the probability of successfully validating counterfeited qtickets grows at most quadratically in the number of such rounds, and hence, 
the likelihood of successful counterfeiting can remain exponentially small even for polynomially large numbers of verifications.

\begin{figure} \includegraphics[width=3.4in]{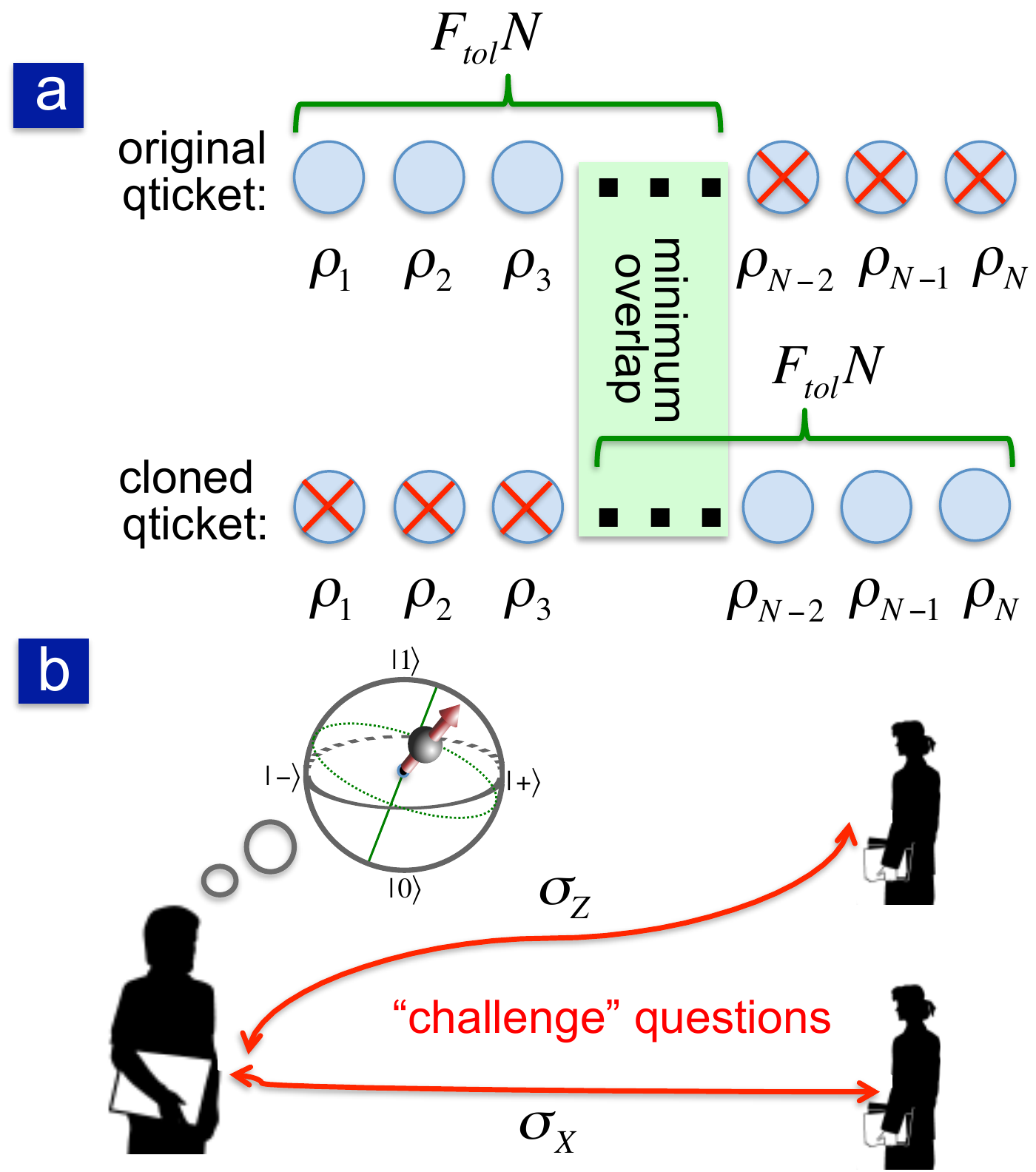} \caption{ a) Depicts the pigeonhole type argument which is utilized in the proof of qticket soundness. For a tolerance fidelity $F_{\tmop{tol}}$, a qticket is only successfully authenticated if it contains at least $F_{\tmop{tol}}N$ valid qubits. However, for two counterfeit qtickets, not all valid qubits must coincide. The minimum number of perfectly cloned qubits enabling both qtickets to be accepted is, $(2F_{\tmop{tol}}-1)N$. b) Depicts the quantum retrieval type situation envisioned for cv-qtickets. For two verifiers asking complementary ``challenge'' questions, the optimal strategy is for the user to measure in an intermediate basis. Such a strategy saturates the tolerance threshold, $ F_{\tmop{tol}}^{\tmop{cv}}=\frac{1+1/\sqrt{2}}{2}$. } \label{fig:PigeonprinPLUScvqticket} \end{figure}

Our previous discussion of qtickets assumed that such tokens are physically transferable to trusted verifiers (e.g. concert tickets); however, in many situations, this assumption of physical deposition, may either be impossible or undesirable.
Recently,  it has been shown \cite{gavinsky_quantum_2011} that it remains possible, even remotely, for a holder to prove the validity of a token by responding to a set of ``challenge'' questions; these questions can only be successfully answered by measuring an authentic token.
The core behind this approach is to ensure that the ``challenge'' questions reveal no additional information about the quantum state of the token.

We now discuss a specific realization of such an approach, the classical verification quantum ticket (cv-qticket), and demonstrate its robustness against noise and operational imperfections.
In contrast to the case of bare qtickets, a cv-qticket holder will be expected to answer ``challenge'' questions and hence to measure qubits himself. One might imagine that the ability to participate in multiple remote verifications simultaneously could offer a dishonest user an additional advantage over the bare qticket case; in particular, certain measurement strategies may yield an increased likelihood for multiple successful authentications.

One example of a cv-qticket framework, is to utilize a set of eight possible two-qubit product states with each qubit prepared along either $X$ or $Z$ (note that a single qubit framework is also possible):
\[
\{ \vert 0, + \rangle, \vert 0, - \rangle, \vert 1, + \rangle, \vert 1,- \rangle,
\vert +, 0 \rangle, \vert -, 0 \rangle, \vert +, 1 \rangle, \vert -,1 \rangle \}.
\]
We then envision each cv-qticket to consist of $n$ blocks, each containing $r$ qubit pairs, and thus, a total of $n\times r\times2$ qubits; as before, each of the qubit pairs is chosen uniformly at random from the allowed set above.
A ``challenge'' question consists of randomly asking the holder to measure each block (of qubits) along either the $X$ or $Z$ basis; naturally, as depicted in Table 1, a valid qubit pair (within a block) is one in which the holder correctly answers the state for the particular qubit (within the pair) which was prepared along the questioned basis. 
For a given tolerance threshold, an overall answer will only be deemed correct if at least $F_{\tmop{tol}}^{\tmop{cv}}r$ qubits within each of the $n$ blocks are found valid. 
By analogy to the qticket case, honest users are exponentially likely to be verified so long as $F_{\tmop{exp}} > F_{\tmop{tol}}^{\tmop{cv}}$; in particular, since there now exist $n$ blocks of qubits, each of which can be thought of as an individual qticket (with $r$ qubits),
 \begin{equation}\label{eq:SoundnessCVqticket}
 p_{\tmop{h}}^{\tmop{cv}} \geq \left(  1- e^{-r D( F_{\tmop{exp}} \| F_{\tmop{tol}}^{\tmop{cv}} )}\right)^n.
\end{equation}
The proof of cv-qticket security is based upon a generalized formalism of quantum retrieval games \cite{gavinsky_quantum_2011, gutoski_toward_2007}, in combination with a generalized Chernoff-Hoeffding bound \cite{impagliazzo_constructive_2010} (details in Supplementary Information).  
So long as  $F_{\tmop{tol}}^{\tmop{cv}} > \frac{1+1/\sqrt{2}}{2}$, a dishonest user is exponentially unlikely to be authenticated by two independent verifiers. 
For two complementary ``challenge'' questions, one finds that on average, no more than $1 +1/\sqrt{2} \approx 1.707$ can be answered correctly. 
Interestingly, the threshold $F_{\tmop{tol}}^{\tmop{cv}}$ corresponds exactly to that achievable by either covariant qubit cloning\cite{bru_phase-covariant_2000} or by measurement in an intermediate basis (Fig. 1b.), suggesting that both such strategies may be optimal \cite{gisin_quantum_2002}.  
Similar to the qticket case, one finds that a dishonest user is exponentially likely to fail,
\begin{equation}\label{eq:SecurityCVqticket}
 p_{\tmop{d}}^{\tmop{cv}} \leq \binom{v}{2}^2 \left( 1/2+e^{-r D( F_{\tmop{tol}} \| \frac{1+1/\sqrt{2}}{2} )} \right)^n,
\end{equation}
where $v$ represents the number of repeated verification attempts (for details see Supplementary Information). Moreover, so long as two verifiers agree to ask complementary ``challenge'' questions, participation in simultaneous verifications is unable to improve a dishonest user's emulated fidelity. 
Thus, in the case of both qtickets and cv-qtickets, so long as the hierarchy of fidelities is such that: $F_{\tmop{dishonest}} < F_{\tmop{tol}} < F_{\tmop{exp}}$, it is possible to prove both soundness and security of each protocol.

\begin{table}[tbp] \centering
\begin{tabular}[c]{|l|cccc|cccc|}
\hline 
\text{Prepare}   & $\vert -,0\rangle$  &$\vert 0,+ \rangle$ &$\vert 1,+ \rangle$ &$\vert 0,+\rangle$ &$\vert 0,+ \rangle$ &$\vert +,1\rangle$ &$\vert -,0\rangle$ &$\vert 1,+ \rangle$\\
\hline
\hline
\text{B:Ask}  & \multicolumn{4}{|c|}{ $Z$ }  & \multicolumn{4}{|c|}{$X$}  \\
\hline
\text{H:Ans.}  & $0,0$  &$0,1$ &$1,1$ &$0,1$ &$-,+$ &$+,-$ &$-,+$ &$+,-$ \\
\hline
\text{Correct}& \OK    &\OK   &\OK   &\OK   &\OK   & \OK  & \OK  & $\times $  \\
\text{Block}& \multicolumn{4}{|c|}{\OK}    & \multicolumn{4}{|c|}{\OK}  \\
\hline
\text{B:Res.} & \multicolumn{8}{|c|}{ Verified } \\
\hline
\end{tabular}
\caption{
Verification of a single cv-qticket. 
Here, we consider a cv-qticket with $n=4$ and $r=2$, totaling $8$ qubit pairs and $F_{\tmop{tol}}= 3/4$ (for illustrative purposes only).
The prepared qubit-pairs are chosen at random, as are the bank's requested measurement bases (for each block).
The holder's answer has at most, a single error per block, which according to, $F_{\tmop{tol}}= 3/4$, is allowed.
Secure cv-qtickets require $F_{\tmop{tol}}>1/2+1/\sqrt{8}$ and a larger number of constituent qubits.
}\label{tab:RegisterVerification}%
\end{table}%


Next, we consider applications of the above primitives to practically relevant protocols.
For instance, one might imagine a composite cv-qticket which allows for multiple verification rounds while also ensuring that the token cannot be split into two independently valid subparts \cite{gavinsky_quantum_2011}. 
Such a construction may be used to create a quantum-protected credit card. Indeed, the classical communication which takes place with the issuer (bank) to verify the cv-qticket (via ``challenge'' questions) may be intentionally publicized to a merchant who needs to be convinced of the card's validity. By contrast to modern credit card implementations, such a quantum credit card would be unforgeable and hence immune to fraudulent charges (Fig. 2a). 

An alternate advantage offered by the qticket framework is evinced in the case where verifiers may not possess a secure communication channel with each other. 
Consider for example, a dishonest user who seeks to copy multiple concert tickets, enabling his friends to enter at different checkpoint gates. A classical solution would involve gate verifiers communicating amongst one another to ensure that each ticket serial number is only allowed entry a single time; however, as shown in Fig. 2b., such a safeguard can be overcome in the event that communication has been severed. By contrast, a concert ticket based upon the proposed qticket primitive would be automatically secure against such a scenario; indeed, the security of qtickets is guaranteed even when verifiers are assumed to be isolated. Such isolation may be especially useful for applications involving quantum identification tokens, where multiple verifiers may exist who are either unable or unwilling to communicate with one another. 

\begin{figure} \includegraphics[width=3.4in]{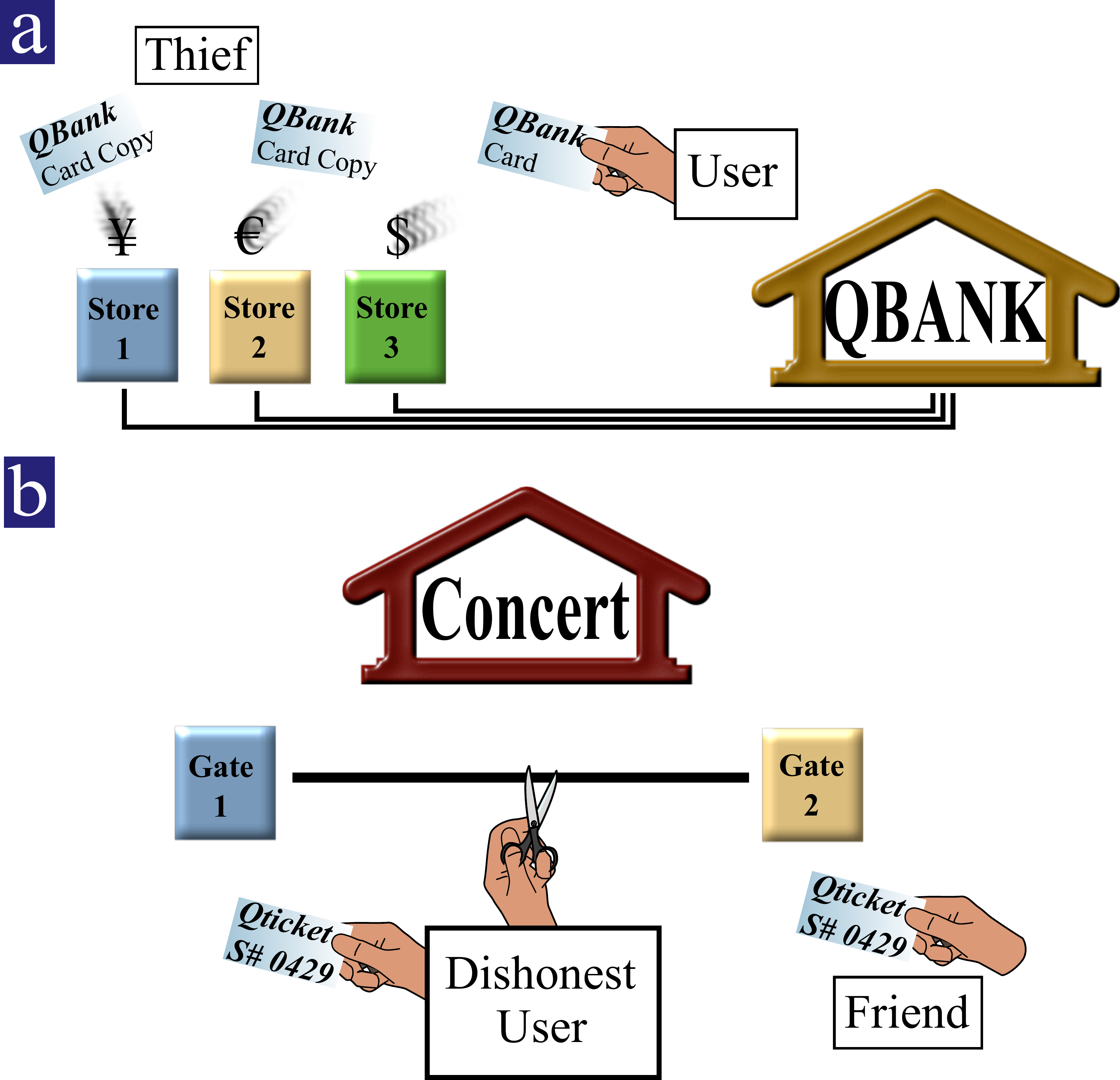} \caption{a) Depicts the possibility of using the cv-qticket framework to implement a quantum-protected credit card. Unlike its classical counterpart, the quantum credit card would naturally be unforgeable; this prevents thieves from being able to simply copy credit card information and perform remote purchases. b) Depicts a dishonest user who attempts to copy a concert qticket (e.g. same serial number), enabling his friend to enter at an alternate checkpoint gate.  Naively, each verifier can communicate with one another to prevent such abusive ticket cloning. However, such a safeguard can be overcome in the event that the communication among verifiers is either unsecured, unavailable or severed (possibly by the dishonest user himself).  The qticket is exempt from this type of attack since security is guaranteed even in the case of isolated verifiers.} \label{fig:MultiBuy} \end{figure}

While quantum primitives have been the subject of tremendous theoretical interest, their practical realization demands robustness in the face of realistic imperfections. Our above analysis demonstrates that such noise tolerance can be achieved for certain classes of unforgeable quantum tokens.  Moreover, the derived tolerance thresholds are remarkably mild and suggest that proof of principle experiments are currently accessible in systems ranging from trapped ions \cite{Hume07, Langer05} and superconducting devices \cite{Wendin03, gladchenko_superconducting_2009} to solid-state spins \cite{Dutt07,Morton08, Balasubramanian09, Maurer11}. In particular, recent advances on single nuclear spins situated in a compact room-temperature solid, have demonstrated that ultra-long storage times can be attained in combination with high fidelity initialization and readout \cite{Maurer11}; such advances suggest that quantum devices based upon single qubit quantum memories may be both practical and realistically feasible.

While our analysis has focused on describing a primitive based upon single tokens, natural extensions to the case of multiple identical quantum tokens open up the possibility of even more novel applications.  In particular, as detailed in the Supplementary Information, it is possible to extend our threshold results to the case where $c$ identical copies of the quantum token are issued. In this case, to ensure that the production of $c+1$ valid tokens is exponentially improbable, the required threshold fidelity must be greater than  $1-\frac{1}{(c+1)(c+2)}$. The existence of such multiple identical tokens can provide a certain degree of anonymity for users and could be employed in primitives such as quantum voting. A crucial question that remains is whether a rigorous proof of anonymity can be obtained in a noisy environment. Furthermore, our proposed quantum tokens can also be seen as a basic noise tolerant building block for implementing more advanced application schemes; such schemes can range from novel implementations of quantum key distribution \cite{bennet_quantum_1984,gisin_quantum_2002,gottesman_proof_2003, scarani_quantum_2008} based upon physical qubit transport to complex one-time-entry identification cards.  
Beyond these specific applications, a number of scientific avenues can be explored, including for example, understanding whether an interplay between computational assumptions and quantum memories can yield fundamentally new approaches to encryption.

\section*{Methods}
{\bf{Proof of Quantum Ticket Soundness}}--- To demonstrate the soundness and security of qtickets, we employ the framework of a Chernoff Bound, which characterizes the central limiting behavior of a set of independent random variables; in particular, it provides exponentially decreasing bounds on tail distributions of their sums. Here, we state a generalization of this bound \cite{impagliazzo_constructive_2010},
\vspace{3mm}

\noindent {\bf Theorem: Generalized Chernoff-Hoeffding bound} 
\noindent{\it Let $X_1, \ldots, X_n$ be Boolean $\{0,1\}$ random variables, such that for some $\delta_i$ and every $S \subseteq \{1,\ldots,n\}$, it holds that $\Pr\left[\bigwedge_{i\in S} X_i\right]\leq \prod_{i\in S}\delta_i$. Then 
\[
\Pr[\sum_{i =1}^n X_i \geq \gamma n] \leq e^{-nD(\gamma \| \delta)}
\]
with $\delta:=n^{-1}\sum_{i=1}^N \delta_i$ and for any $\gamma$ s.t. $\delta\leq \gamma \leq 1$. }
\vspace{3mm}

\noindent $D(p \| q)= p \ln \frac{p}{q} + (1-p) \ln \frac{1-p}{1-q}$ characterizes the distinguishability of two binary probability distributions, where $\text{Pr}(X=1) = p$ for the first distribution and $\text{Pr}(X=1) = q$ for the second. To establish the soundness of qtickets, we now define the ``acceptor'', $P_{\tmop{acc}}^{\rho}$, which projects a pure $N$ qubit product state onto the subspace of valid qtickets. As expected, the size of this subspace will depend on the verifiers tolerance fidelity, $F_{\tmop{tol}}$.
\vspace{3mm}

\noindent {\bf Definition: Acceptance Projector} 

\noindent{\it Given $\rho=\bigotimes_{i=1}^N \rho_i$ and a security parameter $0 \leq F_{\tmop{tol}} \leq 1$, the acceptance projector is given by
\[
P_{\tmop{acc}}^{\rho} = \sum_{\vec{b} : |\vec{b}|_1\geq F_{\tmop{tol}}N} \bigotimes_{i=1}^N \left( b_i \rho_i + \bar{b_i}\rho_i^{\perp} \right) .
\] }
\noindent

\noindent $\vec{b}\in \{0,1\}^{N}$ is a length $N$, boolean string with $|\vec{b}|_1=\sum_{i=1}^N b_i$,  $\bar{b_i} =1 -b_i$, and $\rho_i^{\perp} =\mathbbm{1}-\rho_i $. Intuitively, $|\vec{b}|_1$ can be thought of as a Hamming weight since it characterizes the number of non-zero entries of the string. The sum is over all strings which have at least $F_{\tmop{tol}}N$ entries which are $1$; thus, the definition of $\vec{b}$ naturally  enforces the projection onto the set of valid qticket states. We now recall the qubit map $M_i$ which characterizes the overall fidelity, $F_i$, of encoding, storage and decoding. We define $F_{\tmop{exp}} = 1/N \sum_i F_i$ to be the average achievable experimental fidelity. This brings us to the statement of qticket soundness. 
\vspace{3mm}

\noindent {\bf Theorem: Soundness of a Quantum Ticket} 

\noindent{\it For $F_{\tmop{exp}}>F_{\tmop{tol}}$, an honest holder successfully redeem a qticket with probability
\[
  p_{\tmop{v}}\geq 1 - e^{-N D( F_{\tmop{tol}} \| F_{\tmop{exp}} )}.
\]}
\begin{proof}
 Consider a vector composed of boolean random variables, $\vec{X}=(X_1, \ldots, X_N)$, with a joint probability distribution given by
 \begin{align}
 \Pr[\vec{X} =\vec{b}] &= \frac{1}{|Q|}\sum_{\rho \in Q} \trace\left[  M(\rho) \bigotimes_{i=1}^{N} \left( b_i \rho_i + \bar{b}_i \rho_i^{\perp} \right)\right] \nonumber \\ 
 &= \prod_{i=1}^N \frac{1}{6} \sum_{\rho_i \in \tilde{Q}} \trace\left[ M_i(\rho_i) (b_i\rho_i + \bar{b}_i \rho_i^\perp)  \right] \nonumber
 \end{align}
As evidenced, we can consider $X_i$ to be independent boolean random variables with probability $\Pr[X_i]=F_i$. Moreover, a simple calculation reveals that Eq.~{\bf 1} of the main text can be recast as, $\frac{1}{|Q|}\sum_{\rho \in Q}\trace[P_{\tmop{acc}}^{\rho}M(\rho)] = \Pr[ \sum_{i=1}^N X_i \geq F_{\tmop{tol}}N ]$. Application of the Chernoff bound yields the desired result.
\end{proof}
\vspace{3mm}

\noindent The security proof for qtickets, detailed in the supplementary information, follows in a similar fashion; it requires the generalized Chernoff-Hoeffding bound to rigorously deal with arbitrary counterfeiting attacks, which may in principle generate correlations between qticket components. 

\section*{Acknowledgements}

We thank Y. Chu, C. R. Laumann and S. D. Bennett for insights and discussions. This work was supported in part by the DFG (SFB 631), the QCCC elite network Bayern,  the EU project MALICIA and Catalunya Caixa, the NSF, CUA, DOE (FG02-97ER25308), DARPA (QuEST), MURI, Packard Foundation and the Sherman Fairchild Foundation. 
\bibliography{QuantumMoneyNPhysv22}

\end{document}


\date{\today}

\title{SUPPLEMENTARY INFORMATION \\
Unforgeable Noise-Tolerant Quantum Tokens}

\author{F. Pastawski$^{1}$, N. Y. Yao$^{2}$, L. Jiang$^{3}$, M. D. Lukin$^{2}$, J. I. Cirac$^{1}$}

\affiliation{$^{1}$Max-Planck-Institut fur Quantenoptik, Hans-Kopfermann-Strase 1, Garching, D-85748, Germany}
\affiliation{$^{2}$Physics Department, Harvard University, Cambridge, MA 02138}
\affiliation{$^{3}$Institute for Quantum Information, California Institute of Technology, Pasadena, CA 91125}
\maketitle

\section{Notation and external results}

The following definitions and external results will be used extensively throughout the proofs and are included here to provide a self-contained presentation.

\begin{definition}\label{def:t-design}
A quantum state $t$-design is a probability distribution over pure quantum states $(p_i, \mid \psi_i \rangle)$ such that
\[
  \sum_i p_i \left(\mid \psi_i \rangle \langle \psi_i \mid \right)^{\otimes t} = 
  \int_{\tmop{Haar}}  \left(\mid \psi \rangle \langle \psi \mid \right)^{\otimes t} d \psi.
\]
\end{definition}
In other words, a quantum state $t$-design duplicates the properties of the unique unitarily invariant Haar measure over quantum states for all polynomials up to degree $t$. 
We will use the set of states 
\begin{equation}\label{eq:3design}
 \tilde{Q}=\{ \mid 0 \rangle, \mid 1 \rangle, \mid + \rangle,\mid - \rangle,\mid +i \rangle,\mid -i \rangle\}
\end{equation}
with equal weights $p_i=1/6$; this constitutes a quantum state $3$-design over ${\mathcal H}_2$ \cite{zhu_quantum_2011}.

The average fidelity for a channel quantifies how well the channel preserves quantum states.
\begin{definition}\label{def:avg-fidelity} The Average fidelity of a map $M$ is defined as
\[
 F(M) = \int_{\tmop{Haar}}  \langle \psi \mid M\left( \mid\psi\rangle\langle\psi\mid \right)\mid\psi \rangle d \psi.
\]
\end{definition}
Note for example that the average fidelity of a map $M$ is expressed as a Haar integral of a degree $2$ polynomial expression in bras and kets and can thus be equated to a weighted average over a quantum state $2$-design.

Throughout the text, boolean values ${\mathcal B}=\{True, False\}$ will be represented as $True := 1$, $False := 0$ and the negation $\bar{b} := 1-b$.
We will also use the variable $\vec{b}$ to denote boolean strings (i.e. ordered sequences of values in $\{0,1\}$) with $\tmop{len}(\vec{b})$ denoting the length or number of components of a sequence and $\tmop{tl}(\vec{b})$ denoting the string obtained from removing the last element from $\vec{b}$. We will denote by $\Pr[e]$ the probaility of an event $e$ and $\Exp[v]$ the expectation value of an expression $v$.
Note that according to our convention, if the expression is a boolean formula they may be used interchangeably.

The relative entropy is a distinguishability measure between two probability distributions.
It will be used extensively (particularly among binary or Bernoulli distributions) and appears in the definition of auxiliary results.
Let $0\leq p, q \leq 1$, by abuse of notation, we take $D(p \| q)= p \ln \frac{p}{q} + (1-p) \ln \frac{1-p}{1-q}$, the relative entropy between two Bernoulli probability distributions with respective parameters $p$ and $q$.
Note that this definition satisfies $D(p \| q)\geq 2(p-q)^2$. 

The following generalization of the Chernoff-Hoeffding bound derived by Panconesi and Srinivasan \cite{panconesi_randomized_1997} provides the same thesis as a standard Chernoff bound while relaxing the hypothesis to allow dependent random variables.
\begin{theorem} \label{thm:GeneralizedChernoffBound} (Generalized Chernoff-Hoeffding bound) Let $X_1, \ldots, X_n$ be Boolean $\{0,1\}$ random variables, such that for some $\delta_i$ and every $S \subseteq \{1,\ldots,n\}$, it holds that $\Pr\left[\bigwedge_{i\in S} X_i\right]\leq \prod_{i\in S}\delta_i$. Then 
\[
\Pr[\sum_{i =1}^n X_i \geq \gamma n] \leq e^{-nD(\gamma \| \delta)}
\]
with $\delta:=n^{-1}\sum_{i=1}^N \delta_i$ and for any $\gamma$ s.t. $\delta\leq \gamma \leq 1$. 
\end{theorem}

A further generalization to real valued random variables will also be required. 
This is adapted to our purpose from theorem 3.3 of Impagliazzo and Kabanets \cite{impagliazzo_constructive_2010}.
\begin{theorem}\label{thm:NonBinaryGeneralizedChernoffBound} Let $X_1,\ldots,X_n$ be real valued random variables, with each $X_i\in[0,1]$.
Suppose that there is a $0\leq \delta \leq 1$ s.t., for every set $S\subseteq \{1,\ldots,n\}$,
$\Exp\left[  \prod_{i\in S} X_i\right] \leq \delta^{|S|}$ and $\gamma$ s.t. $\delta\leq\gamma\leq 1$ and $\gamma n \in \mathbbm{N}$. 
Then we have that $\Pr\left[ \sum_{i=1}^{n} X_i \geq \gamma n \right]\leq 2 e^{-nD(\gamma \| \delta)}$.
\end{theorem}

\section{Qtickets}

We first provide a rigorous definition of qtickets and how they are verified.
We then proceed to our claims,  and the soundness, security and tightness of our security bound (accompanied with respective proofs).
Namely, we show that qtickets may be successfully redeemed by an honest holder achieving a sufficiently good storage fidelity.
We then show that a dishonest holder will have a negligible chance of producing two qtickets which are accepted by verifiers from a single valid qticket, even after repeated verification attempts.
Finally we show how a simple counterfeiting strategy has a high probability of producing two such qtickets if the verification tolerance is set below the threshold value.
As an extension, we consider how our results generalize to producing multiple identical qtickets.

\subsection{Definition of qtickets}

Each qticket consists of a serial number $s$ and an $N$ component pure product state $\rho^{(s)} = \bigotimes_{i=1}^N \rho_i^{(s)}$. 
For each serial number $s$, qticket components $\rho^{(s)}_i$  are chosen uniformly at random from $\tilde{Q}$.
This means qtickets $\rho^{(s)}$ are taken uniformly at random from the set $Q=\tilde{Q}^{\otimes N}$ 
(where by abuse of notation, the elements of $Q$ are $N$ component pure product states  in $\mathcal{H}_Q = \mathcal{H}_2^N$, with components taken from $\tilde{Q}$).
The verifiers store a database containing, for each $s$, a classical description of $\rho^{(s)}$  kept secret from ticket holders and the general public.
In order to simplify notation, the serial number $s$ associated to individual qtickets will be omitted from now on.

In order to use qtickets, they are transferred to a verification authority who can either accept or reject them.
In both cases however, the qticket is not returned, only the binary outcome of verification.
The qticket protocol is additionally parametrized by the fraction $F_{\tmop{tol}}$ of qubits that a verification authority requires to be correct in order for verification to succeed.
In order to verify a submitted qticket $\tilde{\rho}$, a full measurement will be performed in the product basis associated to the original qticket $\rho$ and the number of correct outcomes is then counted.
If more than at least $F_{\tmop{tol}}N$ are correct, the (possibly noisy) submitted qticket $\tilde{\rho}$ is accepted, otherwise, it is rejected.

For any pure product state $\rho=\bigotimes_{i=1}^N \rho_i$ we define a projector $P_{\tmop{acc}}^{\rho} \in {\mathcal L}( {\mathcal H}_Q )$ associated to the subspace of states that would be accepted if $\rho$ were a qticket (i.e. states coinciding with $\rho$ in at least a fraction $F_{\tmop{tol}}$ of the qubits). 
The projector $P_{\tmop{acc}}^{\rho}$ offers a more abstract interpretation and may be rigorously defined as 
\begin{definition}[Acceptance projector]\label{def:AcceptanceProjector}
Given a pure $N$ qubit product state $\rho=\bigotimes_{i=1}^N \rho_i$ and a security parameter $0 \leq F_{tol} \leq 1$, we define the acceptance projector
\[
P_{\tmop{acc}}^{\rho} = \sum_{\vec{b} : \sum b_i \geq F_{tol}N} \bigotimes_{i=1}^N \left( b_i \rho_i + \bar{b_i}\rho_i^{\perp} \right),
\]
where $\vec{b}\in \{0,1\}^{N}$ is a boolean string.
\end{definition}
By abused of notation, $\rho_i$ and its orthogonal complement  $\rho_i^\perp := \mathbbm{1}_2-\rho_i$ are used as  rank 1 projectors in ${\mathcal L}({\mathcal H}_2)$.

\subsection{Soundness}

The soundness result states that even under imperfect storage and readout fidelity, legitimate qtickets work well as long as the fidelity loss is not too severe.
The completely positive trace preserving (CPTP) maps $M_i$ will be assumed to represent the encoding, storage and readout of the $i$-th qubit component of the qticket.
In this sense, the soundness statement takes place at the level of single qubits.
This is necessarily the case, since legitimate qtickets are ruined if a significant fraction of the qubits fail in a correlated way.
Given $F_i = F(M_i)$, the average fidelity of the qubit map $M_i$, we define $F_{\tmop{exp}}:=N^{-1}\sum F_{i}$ to be the average qubit fidelity of the full map $M=\bigotimes_i M_i$ over all components.
The probability that the ``noisy'' qticket resulting from this map is accepted as valid is given by
$p_{\tmop{h}}(M)=\frac{1}{|Q|} \sum_{\rho \in Q} \trace\left[  P_{\tmop{acc}}^{\rho}  M\left(  \rho \right)  \right]$.
\begin{theorem}\label{Theorem:Soundness}
(Soundness of qtickets) As long as $F_{\tmop{exp}}>F_{\tmop{tol}}$, an honest holder can successfully redeem qtickets with a probability
\[
  p_{\tmop{h}}(M) \geq 1 - e^{-N D( F_{\tmop{tol}} \| F_{\tmop{exp}} )}.
\]
\end{theorem}
\begin{proof}
Consider the boolean random variables $\vec{X}=(X_1, \ldots, X_N)$ with joint distribution given by
\begin{equation}
 \Pr[\vec{X}=\vec{b}] = \frac{1}{|Q|}\sum_{\rho \in Q} \trace\left[  M(\rho) \bigotimes_{i=1}^{N} \left( b_i \rho_i + \bar{b}_i \rho_i^{\perp} \right)\right].
\end{equation}
Since $M=\bigotimes_i M_i$, we may recast Eqn. S2 as
\begin{equation}
 \Pr[\vec{X}=\vec{b}] = \prod_{i=1}^N \frac{1}{6} \sum_{\rho_i \in \tilde{Q}} \trace\left[ M_i(\rho_i) (b_i\rho_i + \bar{b}_i \rho_i^\perp)  \right]
\end{equation}
Since $\tilde{Q}$ is a quantum state 2-design over qubit space, each factor coincides with the definition of the average fidelity $F_i$ of $M_i$ if $b_i=1$ and with $1-F_i$ if $b_i=0$.
Hence the $X_i$ are independent boolean random variables with probability $\Pr[X_i]=F_i$.
Moreover, according to definition \ref{def:AcceptanceProjector}, 
we have $\frac{1}{|Q|}\sum_{\rho \in Q}\trace[P_{acc}^{\rho}M(\rho)] = \Pr[ \sum_{i=1}^N X_i \geq F_{\tmop{tol}}N ]$.
Since the $X_i$ are independent, a standard Chernoff-Hoeffding bound allows us to conclude.
\end{proof}

\subsection{Security}

The security statement expresses that for a sufficiently large security parameter $F_{\tmop{tol}} > 5/6$, the average probability of a dishonest participant successfully forging two accepted
qtickets from a single one is exponentially small in $N$.
Indeed, when simultaneously submitting two qtickets produced by applying the most general possible transformation $T$ (a CPTP map) on a single valid qticket, the probability of both of them being accepted is exponentially small in $N$ as long as we are not given access to the classical description.
\begin{definition}\label{Def:CounterfeitingFidelity}(Counterfeiting fidelity)
We define the average counterfeiting fidelity of a map $T \in {\mathcal H}_Q\rightarrow {\mathcal H}_Q^{\otimes 2}$ as
\begin{equation}
  p_{\tmop{d}}(T) = 
  \frac{1}{|Q|} \sum_{\rho \in Q} \trace\left[  \left( P_{\tmop{acc}}^{\rho}  \right)^{\otimes2}  T\left(  \rho \right)
\right]
\end{equation}
\end{definition}
The following is one of the main results and states that as long as the verification threshold $F_{\tmop{tol}}$ is set sufficiently high a counterfeiter will have negligible chances of producing two valid qtickets from a single genuine original.
\begin{theorem}\label{Theorem:QticketSecurity}
(Security of qtickets) For $F_{\tmop{tol}} > 5/6$ and for any CPTP map $T \in {\mathcal H}_Q\rightarrow {\mathcal H}_Q^{\otimes 2}$ we have that
\begin{equation}
 p_{\tmop{d}}(T) \leq e^{-N D( 2F_{\tmop{tol}}-1 \| 2/3 )}.
\end{equation}
\end{theorem}
Most of the work for proving this goes into excluding the possibility that a non-product counterfeiting strategy could perform significantly better than any product strategy such as performing optimal cloning on each individual qubit.
That is, we take into account the fact that the map $T$ need not factorize with respect to the different components of the qticket.
Note also that $D( 2F_{\tmop{tol}}-1 \| 2/3 )=0$ precisely for $F_{\tmop{tol}}=5/6$ and is positive otherwise.
Finally, we prove that even if the holder of a qticket attempts to perform $v$ succesive verification attempts (each time possibly using information learned from the acceptance/rejection of previous attempts) the chances of having two or more submitted qtickets accepted grows by no more than a factor of $\binom{v}{2}$. 
\begin{theorem}\label{Theorem:SecuriyWithLearning}(Security of qtickets with learning) If the holder of a valid qticket submits $v$ tokens for verification, the probability of having two or more accepted is upper bounded by 
\[
p_{\tmop{d},v} = \binom{v}{2}e^{-N D( 2F_{\tmop{tol}}-1 \| 2/3 )}.
\]
\end{theorem}
Note that since $\binom{v}{2}$ is a polynomial of degree $2$ in $v$, this bound still allows for an exponentially large number (in $N$) of qticket submissions $v$, while preserving exponentially good security.

\subsubsection{Proof outline}

We now outline the proof for theorems \ref{Theorem:QticketSecurity} and  \ref{Theorem:SecuriyWithLearning}.
First, the claim in theorem \ref{Theorem:QticketSecurity} is equated to an equivalent one, which averages over the set of all pure product states instead of $Q$.
We then bound the average cloning probability by $(2/3)^N$ for the set of pure product states following the lines of R. F. Werner \cite{werner_optimal_1998} for the optimal cloning of pure states.
From there, the generalized Chernoff bound from theorem \ref{thm:GeneralizedChernoffBound} for dependent random variables allows us to derive the desired result.
The result of theorem \ref{Theorem:SecuriyWithLearning} is obtained from a counting argument relating the security of multiple verification attempts with the static counterfeiting fidelity bound of theorem \ref{Theorem:QticketSecurity}.

\subsubsection{Equivalence with continuous statement}

For the qticket protocol, drawing each component from a discrete set of states is required in order to provide an efficient classical description.
However, certain statements are simpler to analyze over the full set of pure product states.
This is the case for the counterfeiting fidelity, which can also be expressed as a uniform average over all pure product states.
\begin{lemma}\label{Lemma:CounterfeitingFidelity}(Counterfeiting fidelity)
The average counterfeiting fidelity of a map $T$ can be expressed as
\begin{equation}
  p_{\tmop{d}}(T) = 
  \int d\vec{\rho} \quad \trace\left[  \left( P_{\tmop{acc}}^{\vec{\rho}}  \right)^{\otimes2}  T\left(  \vec{\rho} \right)
\right]
\end{equation}
where $\int d \vec{\rho}$ represents $N$ nested integrations on the Haar measure of qubit components and $\vec{\rho}$ the resulting product state.
\end{lemma}
\begin{proof}
Definition \ref{def:AcceptanceProjector} may seem unnecessarily cumbersome, yet it serves to make explicit that the projector $P_{\tmop{acc}}^{\rho}$ is a polynomial of degree 1 in each of the components $\rho_i$ of the qticket $\rho$.
Furthermore, note that regardless of what the multi-qubit map $T$ is, its application $T(\rho)$ has degree 1 in each of the components $\rho_i$ of $\rho$.
Together this implies that the integrand of lemma \ref{Lemma:CounterfeitingFidelity} is an polynomial of degree at most 3 in each of the qubit components $\rho_i$ of $\vec{\rho}$.
We may conclude by observing that the average taken in definition \ref{Def:CounterfeitingFidelity} is equivalent to uniformly taking each component $\rho_i$ from a qubit state 3-design.
\end{proof}

\subsubsection{Optimal cloning for pure product states}

R. F. Werner \cite{werner_optimal_1998} obtained a tight upper bound for the average probability of a CPTP map $T$ producing $m$ clones from $n$ copies of an unknown pure quantum state $\mid \psi \rangle$.
Our statement is that if one attempts to clone an $N$ component pure product state, the optimal cloning probability is achieved by independently cloning each of the components; neither generating entanglement nor correlations may help with the cloning.
We present this statement for the case of cloning two copies from a qubit product state, but the derivation is fully generalizable.
\begin{lemma}\label{lemma:optimalCloningOfPureProducts}(Optimal cloning of pure product states) The average cloning fidelity over $N$ qubit component pure product states of a CPTP map $T$ is bounded by
\[
\int d\vec{\rho}  \trace[\vec{\rho}^{\otimes 2} T(\vec{\rho})] \leq \left( 2/3\right)^N .
\] 
\end{lemma}
\begin{proof}
One possible derivation of this lemma is by following the lines of the original proof for optimal cloning of pure states \cite{werner_optimal_1998}.
First one shows that if there is a cloning map $T$ achieving average fidelity $F^\star$ then there is a covariant cloning $T^\star$ achieving the same average fidelity.
This map can be explicitly constructed as
\begin{equation}
   T^\star( \vec{\rho} ) = \int d \vec{g} \quad \vec{g}^{\dagger \otimes 2} T( \vec{g} \vec{\rho} \vec{g}^{\dagger}) \vec{g}^{\otimes 2},
\end{equation}
where the integral $\int d \vec{g}$ averages over all possible local rotations $\vec{g}$ on $N$ subsystems.
This covariant map achieves exactly the same cloning fidelity for any initial pure product state since all pure product states are equivalent up to local unitaries.
Finally, we observe
\begin{equation}
 0\leq \trace[\vec{\rho}^{\otimes 2} T^\star({\mathbbm 1}_{2^N} - \vec{\rho})]
\end{equation}
since ${\mathbbm 1}_{2^N} - \vec{\rho}$ is positive and $T^\star$ positivity preserving. We then obtain
\begin{equation}
 F^\star \leq \trace[\vec{\rho}^{\otimes 2} T^\star({\mathbbm 1}_{2^N})]
\end{equation}
and  may now average this inequality over $\vec{\rho}$ and use
\begin{equation}
 \int d\vec{\rho} \quad \vec{\rho}^{\otimes 2} = \frac{(S_2)^{\otimes N}}{3^N},
\end{equation}
where $S_2$ is the rank 3 projector onto the symmetric space of two qubits.
The operator norm of this expression is $1/3^N$ whereas $\trace[T^\star({\mathbbm 1}_{2^N})]\leq 2^N$ leading to $F^\star \leq (\frac{2}{3})^N$, as desired.
\end{proof}

\subsubsection{Pigeonhole argument and Chernoff bound}

We are now ready to prove the first no-counterfeiting result for qtickets.

\begin{proof}[Proof of theorem \ref{Theorem:QticketSecurity}]
Consider the boolean random variables $\vec{E}=(E_1, \ldots, E_N)$ with joint distribution given by
\begin{equation}
 \Pr[\vec{E}=\vec{b}] = \int d\vec{\rho} \trace\left[  T(\vec{\rho}) \bigotimes_{i=1}^{N} \left( b_i \rho_i^{\otimes 2} + \bar{b}_i ({\mathbbm 1}_{4}-\rho_i^{\otimes 2}) \right)\right].
\end{equation}
Intuitively, the variable $E_i$ represents the event of measuring the $i$-th component to be correctly cloned.

In order for the two qtickets to be accepted, there must be a total of at least $F_{\tmop{tol}}N$ components yielding the correct measured outcome in each qticket.
By the pigeonhole principle, this means that there are at least $2F_{\tmop{tol}}N-N$ components which were measured correctly on both submitted qtickets,
\begin{equation}\label{eq:Pigeonholebound}
 p_{\tmop{d}} (T) \leq \Pr\left[ \sum_{i=1}^N E_i \geq (2F_{\tmop{tol}}-1)N \right].
\end{equation}
For arbitrarily chosen $T$, the $E_i$ may be dependent variables.
However, according to lemma \ref{lemma:optimalCloningOfPureProducts}, for any subset $S$ of qubit components, we may bound 
\begin{equation}
 \Pr[ \forall_{i\in S}  E_i ] \leq \left( \frac{2}{3} \right)^{|S|}.
\end{equation}
Theorem \ref{thm:GeneralizedChernoffBound}, is now invoked to provide an upper bound on the RHS of eq. \ref{eq:Pigeonholebound}, yielding the thesis of theorem \ref{Theorem:QticketSecurity}.
\end{proof}

\subsubsection{Combinatorial bound on learning}\label{sec:CombinatorialBoundOnLearning}

The bound on counterfeiting that we have provided assumes that two (possibly entangled) counterfeits are produced by applying a CP map on a single original copy.
In contrast, a sequential strategy temporally orders the submitted qtickets where the production strategy (CP map) for the later submissions can depend on whether previous submissions where accepted or not.
The counterfeiter may learn valuable information about how to construct valid qtickets from the feedback provided by the verifiers.
The content of theorem \ref{Theorem:SecuriyWithLearning} is that even with a valid qticket and the information learned from $v$ repeated submissions it is very unlikely for a counterfeiter to produce more than one accepted qticket.

\begin{proof}[Proof of theorem \ref{Theorem:SecuriyWithLearning}]
According to theorem \ref{Theorem:QticketSecurity}, the probability $p_{\tmop{d}}(T)$ for any CP map $T$ to produce two valid counterfeit copies from a single one, is upper bounded by $B= e^{-ND(2F_{\tmop{tol}}-1  \| 2/3)}$.
We bound the counterfeiting probability of an interactive strategy $S$ submitting $v$ tokens for verification by the sum of the counterfeiting fidelity of $\binom{v}{2}$ CP maps $T_{k,l}$.
Each of these maps corresponds to the case in which a specific pair $(k,l)$ of the $v$ submitted tokens are the first to be accepted by the verifiers.

Without loss of generality, we assume that in an interactive strategy the holder waits for the outcome of the $j$-th verification in order to decide how to continue and produce the $j+1$-th submission.
We model a $v$ step interactive strategy $S$ as a collection of CPTP maps $\{ S_{\vec{b}}\}$ with $\vec{b}$ a boolean string of length between 0 and $v-1$ representing what the counterfeiter does after receiving the first $\tmop{len}(\vec{b})$ verification outcomes.

Each $S_{\vec{b}}$ is a CPTP map from ${\mathcal H}_H$ to ${\mathcal H}_Q \otimes {\mathcal H}_H$, where ${\mathcal H}_Q$ is a Hilbert space accommodating qtickets
and ${\mathcal H}_H$ is a larger space representing the memory of the holder.

For any partial verification result $\vec{b}$ we may write the CPTP map which produces the $\tmop{len}(\vec{b})$ submissions as $\tilde{S}_{\tmop{tl}(\vec{b})}$, which is composed of successively applying $S_{\vec{b'}}$ for all initial substrings $\vec{b'}$ of $\vec{b}$. 
That is
\begin{equation}
\begin{split}
 \tilde{S}_\emptyset := & S_{\emptyset} \\
 \tilde{S}_{\vec{b}} := & \left( \tmop{id}_{Q}^{\otimes \tmop{len}(\vec{b}) } \otimes S_{\vec{b}} \right) \circ \tilde{S}_{\tmop{tl}(\vec{b})}.
 \end{split}
\end{equation}
\begin{figure}[phtb]
\includegraphics[width=0.9\columnwidth]{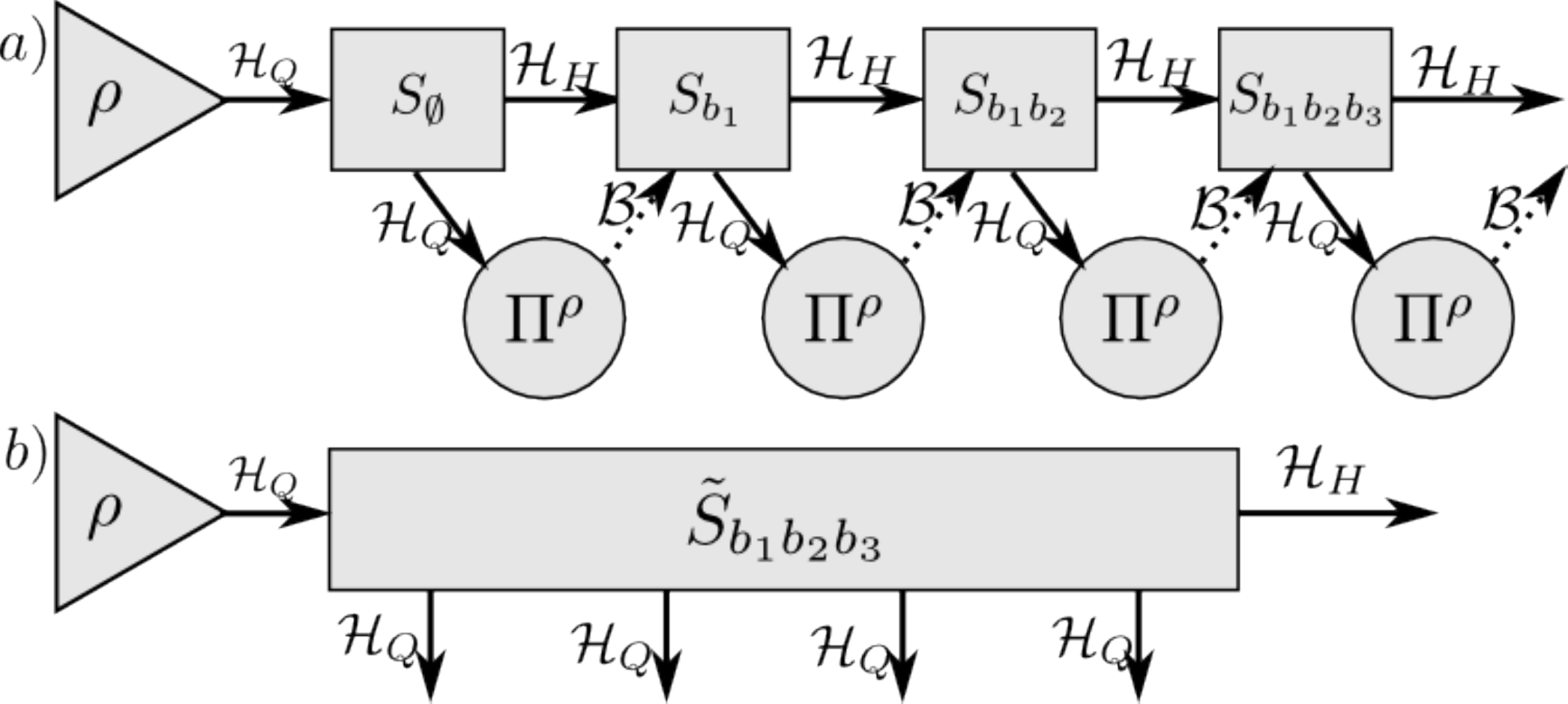}\caption{
a) We schematically illustrate how a dynamical strategy $S$ works.
Each step of a strategy (grey rectangles) is a CPTP map $S_{\vec{b}} $ which depends on the classical outcome $\vec{b}$ of previous verifications.
The first map $S_{\emptyset}$ takes an original qticket $\rho$ as input, whereas subsequent steps rely on an internal memory state of the holder.
The content of internal memory could range from no information at all, to a full original qticket and a detailed register of previous submissions.
The verifiers have a fixed strategy $\Pi^\rho$ which consists of applying the measurement $\{P^\rho_{\tmop{acc}},P^\rho_{\tmop{rej}}\}$ and only returning the classical boolean measurement outcome.
b) By fixing the classical input $\vec{b}$ to the strategy, a CPTP map $\tilde{S}_{\vec{b}} \in {\mathcal H}_Q \rightarrow {\mathcal H}_Q^{\otimes \tmop{len}(\vec{b})+1} \otimes {\mathcal H}_H $ is constructed, corresponding to one possible partial application of the strategy $S$. 
This CPTP map produces $\tmop{len}(\vec{b})+1$ possibly entangled outputs in $\mathcal{H}_Q$ from a single input qticket.
}
\label{fig:Learning1}
\end{figure}

For an interactive strategy $S$ the probability that the first $\tmop{len}(\vec{b})$ verification outcomes are given by $\vec{b}$ is expressed as
\begin{equation}
 p_{\vec{b}}(S)= \frac{1}{|Q|}\sum_{\rho \in Q} \trace[ \tilde{S}_{\tmop{tl}(\vec{b})}(\rho) 
 \bigotimes_{j=1}^{\tmop{len}(\vec{b})} \left( b_j P_{\tmop{acc}}^\rho + \bar{b}_j P_{\tmop{rej}}^\rho  \right) \otimes {\mathbbm 1}_H ],
 \end{equation}
where $P_{\tmop{rej}}^{\rho}:= {\mathbbm 1}_{Q}-P_{\tmop{acc}}^{\rho}$.
The probability for the interactive strategy $S$ to succeed at counterfeiting in $v$ steps can be described as the sum of these probabilities over all possible full verification outcomes including at least two acceptances
\begin{equation}
 p_{\tmop{d},v}(S) = \sum_{\substack{\vec{b}: \sum b_i\geq 2 \\ \tmop{len}(\vec{b})=v }} p_{\vec{b}}(S).
\end{equation}
The key idea now is to use $p_{\vec{b}}(S)=p_{\vec{b}0}(S)+p_{\vec{b}1}(S)$ to provide an alternate expression for this sum.
Namely, we combine verification outcomes starting in the same way into a single summand while avoiding the inclusion of failed counterfeiting attempts.
Each full verification outcome containing two or more successful verifications has a unique shortest initial substring containing exactly two successful verifications.
That a given substring is the shortest can be guaranteed by taking the last verification of the substring to be one of the two accepted.
\begin{equation}\label{eq:LearningAlternateSum}
  p_{\tmop{d},v}(S) = \sum_{\substack{ \vec{b}: \sum b_i = 2 \\ b_{\tmop{len}(\vec{b})}=1} } p_{\vec{b}}(S).
\end{equation}
Each of the $\binom{v}{2}$ summands on the RHS of Eq. (S\ref{eq:LearningAlternateSum}), 
may be characterized by two indices $k,l$ s.t. 
\begin{equation}
\vec{b}=\overbrace{0\ldots0}^{k-1}1\overbrace{0\ldots 0}^{l-k-1}1\quad \text{for some }k<l\leq v .
\end{equation}
For each one of these summands, we construct a static strategy $T_{k,l}(\rho) = \trace_{\backslash k,l}[ \tilde{S}_{\tmop{tl}(\vec{b})}(\rho)]$ which takes as input a single valid qticket $\rho$ and submits exactly two tokens. The counterfeiting probability of this map on $\rho$ is
\begin{equation}
\begin{split}
 & \trace\left[ \left(P_{\tmop{acc}}^\rho\right)^{\otimes 2} T_{k,l}(\rho) \right]\\
 =& \trace\left[ \left(P_{\tmop{acc}}^\rho\right)^{\otimes 2} \trace_{\backslash k,l}[ \tilde{S}_{\tmop{tl}(\vec{b})}(\rho)]\right] \\
=& \trace[ \tilde{S}_{\tmop{tl}(\vec{b})}(\rho) 
\bigotimes_{j=1}^{\tmop{len}(\vec{b})} \left( b_j P_{\tmop{acc}}^\rho + \bar{b}_j {\mathbbm 1}_Q  \right) \otimes {\mathbbm 1}_H ] \\
\geq& \trace[ \tilde{S}_{\tmop{tl}(\vec{b})}(\rho) 
\bigotimes_{j=1}^{\tmop{len}(\vec{b})} \left( b_j P_{\tmop{acc}}^\rho + \bar{b}_j P_{\tmop{rej}}^\rho  \right) \otimes {\mathbbm 1}_H ].
\end{split}
\end{equation}
By averaging over $\rho \in Q$ we obtain $p_{\vec{b}}(S) \leq p_{\tmop{d}}(T_{k,l}) \leq B$ and invoking Eq. (S\ref{eq:LearningAlternateSum}) we obtain $p_{\tmop{d},v}(S) \leq \binom{v}{2} B$.
\end{proof}

\subsection{Tightness}

\begin{figure}[ptb]
\includegraphics[width=1.1\columnwidth]{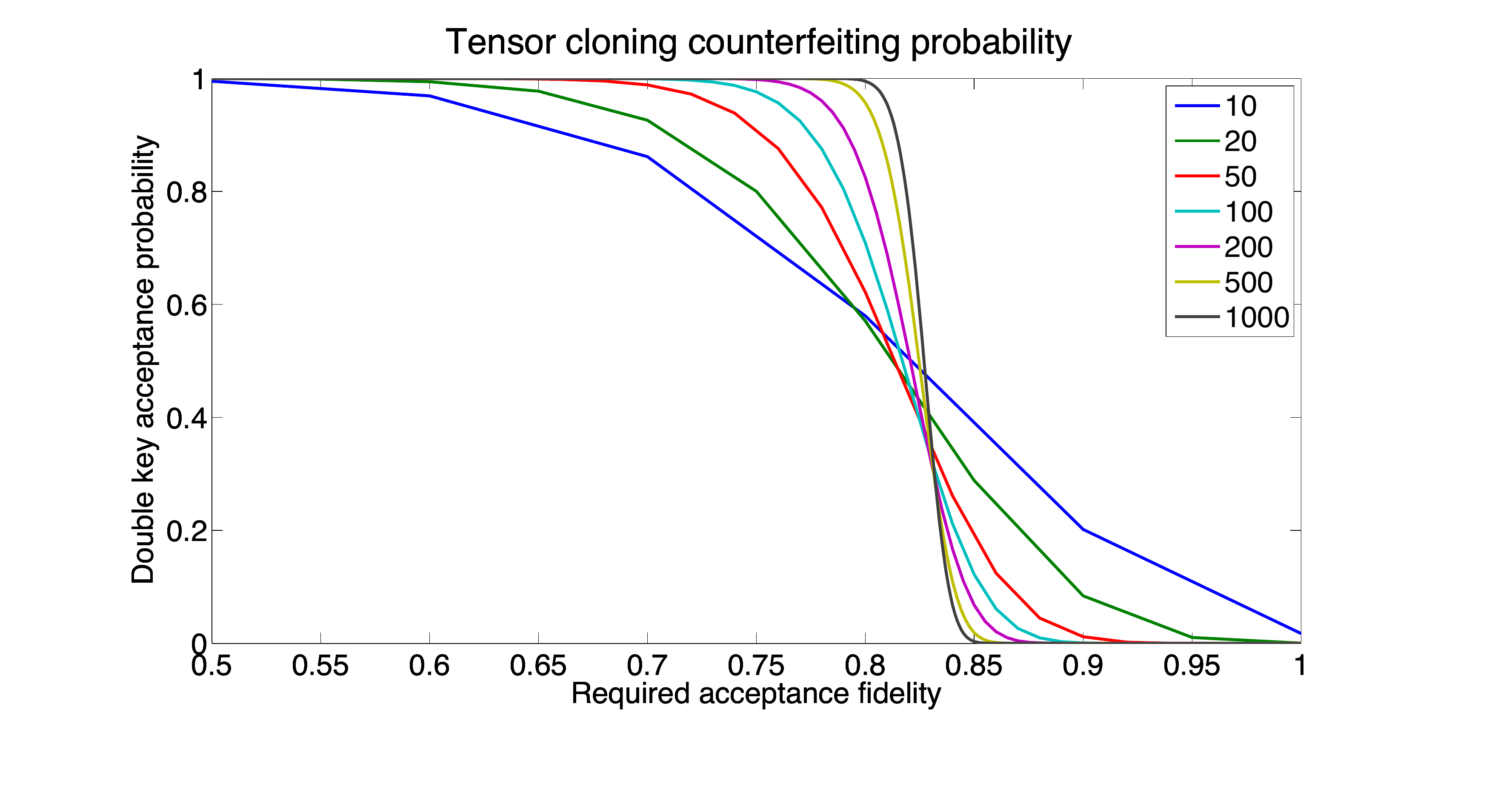}\caption{
We numerically calculate the probability of accepting two copies of a qticket
when the adversary strategy is assumed to be independently cloning each of the
$N$ qubits using an optimal cloning map. 
We see that the probability of producing two accepted qtickets approaches a step function at $5/6$ with $N$.}
\label{fig:DoubleAcceptanceProbability}
\end{figure}

For $F_{tol} < 5/6$ applying an optimal qubit cloning map\cite{werner_optimal_1998} $\Lambda(\rho) = \frac{1}{3} \rho\otimes\rho+ \frac{1}{6} \rho\otimes\mathbbm{1} + \frac{1}{6} \mathbbm{1} \otimes\rho$ on each of the individual 
qubits of the qticket provides a good counterfeiting probability.
The plot in Fig. S\ref{fig:DoubleAcceptanceProbability} illustrates the probability of counterfeiter to actually get two qtickets accepted when taking this approach.
For each of the two counterfeited qtickets, the probability of failing verification is the cumulant of a binomial distribution $B(N, 5/6)$ up to $F_{\tmop{tol}}N$ and rejection probability may be upper bounded by $\frac{1}{2}\exp(-2N(5/6-F_{\tmop{tol}})^2)$ using Hoeffding's inequality.
Even when failure of the two qtickets is anticorrelated, the probability of either of them failing verification can not excede the sum.
This shows that a the scheme can not be made secure for $F_{\tmop{tol}}<5/6$.
While such a scheme provides optimal forging probability when ($F_{tol} = 1$), other
schemes could in principle outperform it in terms of counterfeiting capability.
Althought this is in principle possible, our security result shows that assymptotically in $N$, no other strategy may work for  $F_{\tmop{tol}}>5/6$.

\subsection{Extension: Issuing multiple identical qtickets}

Our results admit generalization to a scenario where the $c$ identical copies of each qtickets are issued and succesful verification of $c+1$ is to be excluded.
To obtain an analog of lemma \ref{Lemma:CounterfeitingFidelity} requires the individual qubits composing a qticket to be drawn at random from a state $t$-design with $t=c+(c+1)$ (for example $t=5$ would already be needed if two identical copies are issued).
The optimal $c\rightarrow c+1$ cloning probability for $N$ component product states is in this case bounded by $\left(\frac{c+1}{c+2}\right)^N$.
The threshold fidelity required to guarante security is then given by $F_{\tmop{tol}} > 1-\frac{1}{(c+1)(c+2)}$
For such an $F_{\tmop{tol}}$, the analogous result to theorem \ref{Theorem:QticketSecurity} one obtained is
\begin{equation}
 p_{c\rightarrow c+1} (T) \leq e^{-N D\left( (c+1)F_{tol}-c \| \frac{c+1}{c+2} \right)}.
\end{equation}
Finally, if $v>c+1$ verification attempts are allowed, the probability of counterfeiting can be proven not to grow faster than $\binom{v}{c+1}$.
The proofs of these claims completely follow the lines that have been presented.
Striving for legibility, we have limited the proof presented to $c=1$, thus avoiding the notational burdon imposed by the extra indices required.

\section{CV-Qtickets}

In this section we provide a proof that cv-qtickets are secure, not only against counterfeiting but also against any other possible double usage.
We first present definitions for cv-qtickets and their verification.
We then state the associated soundness and security guarantees and outline the security proof.
Only the proof of the security statement is provided, since proving soundness for cv-qtickets requires no additional techniques as compared to soundness of qtickets.

\subsection{Definition of cv-qticket}

Each cv-qticket is composed of $n\times r$ qubit pairs.
Each qubit pair is prepared by choosing a state from 
\[
\{ \vert 0, + \rangle, \vert 0, - \rangle, \vert 1, + \rangle, \vert 1,- \rangle,
\vert +, 0 \rangle, \vert -, 0 \rangle, \vert +, 1 \rangle, \vert -,1 \rangle \}
\]
uniformly at random.


A full verification question for the cv-qticket will consist of $n$ randomly chosen axes from $\{X,Z\}$ each corresponding to a specific block of $r$ qubit pairs.
In principle, the holder of the cv-qticket then measures the polarization of every qubit components  along the corresponding requested axis and communicates the measurement outcomes to the verifier.
The criteria to consider an answer correct is the following; within each of the $n$ blocks, at least $F_{\tmop{tol}}r$ of the reported outcomes corresponding to qubits prepared in a polarization eigenstate of the inquired axis should be given correctly.

\subsection{Soundness}

The soundness result states that even under imperfect storage and readout fidelity, legitimate cv-qtickets work well as long as the fidelity loss is not too severe.
Again, the completely positive trace preserving (CPTP) maps $M_j$ will be assumed to represent the encoding, storage and readout of the $j$-th qubit component of the cv-qticket, with the full map over all components given by $M=\bigotimes_{j\in \{1,\ldots,2r\times n\}} M_j$ .
In the case of cv-qtickets, sufficiently many $(F_{\tmop{tol}}r)$ correct answers should be provided within each block, demanding that a sufficiently good average fidelity be implemented for every single block. 
A random remapping of the Cartesian axes for each qubit component of a cv-qticket is also necessary, and can be achieved via a random unitary (possibly from a unitary 2-design).
This is required for example in the case where an actual physical polarization, say $X$, is lost faster than other components.
In this case asking for the stored $X$ polarization for all qubits in a block may yield a large failure probability even though the average storage fidelity among the qubits is sufficiently high.
A random unitary remapping solves this and allows to connect with the average qubit storage fidelity, even in the case where only two nominal axes are used.

Given $F_j = F(M_j)$, the average fidelity of the qubit map $M_j$, we define $F_{\tmop{exp},b}:=N^{-1}\sum_{j:\lceil \frac{j}{2r} \rceil = b} F_{j}$ to be the average qubit fidelity within block $b\in\{1,\ldots,n\}$.
Furthermore, to simplify the final expression, let us define $F_{\tmop{exp}}=\min_b F_{\tmop{exp},b}$.

\begin{theorem}\label{Theorem:CVSoundness}
(Soundness of cv-qtickets) As long as $F_{\tmop{exp}}>F_{\tmop{tol}}$, an honest holder implementing a map $M$ can successfully redeems cv-qtickets with a probability
\[
  p_{\tmop{h}}^{\tmop{cv}}(M) \geq \left( 1 - e^{-r D( F_{\tmop{exp}} \| F_{\tmop{tol}}  )} \right)^n.
\]
\end{theorem}
Observe that one may reduce this statement to $n$ independent statements within each block which are completely analogous to the soundness for qtickets theorem \ref{Theorem:Soundness}.

\subsection{Security}

A naive security statement expresses that the holder of a single cv-qticket is unable to produce two copies from it, each with the potential of passing a verification.
Since the verification of cv-qtickets is achieved by sending a classical message to a verifier, the security statement for cv-qtickets goes beyond this; it states that even with simultaneous access to two randomly chosen verification questions, the holder of a cv-qticket is exponentially unlikely to provide satisfactory answers to both.
We further extend our security claim, to an even more adverse scenario; the holder of a cv-qticket has simultaneous access to $v$ independent verification questions and may proceed to answer them in any chosen order. 
Moreover failing in verification attempts does not forbid the holder from further attempts which may possibly be performed relying on the information accumulated from previous verification outcomes.

Let $S$ be a mathematical object denoting the counterfeiting strategy taken by the holder of a valid cv-qticket.
We will denote by $p_{\tmop{d},v}^{\tmop{cv}}(S)$, the probability that strategy $S$ leads to two or more successful verifications when engaging in $v$ verification attempts with possibly independent verifiers.
The probability is taken over the random generation of cv-qtickets, of verification questions, and of measurement outcomes (Born's rule).
The security statement is then
\begin{theorem}\label{Thm:CVSecurity}(Security of cv-qtickets) For any counterfeiting strategy $S$ and tolerance parameter $F_{\tmop{tol}}>\frac{1+1/\sqrt{2}}{2}$ we have 
\[
  p_{\tmop{d},v}^{\tmop{cv}}(S) \leq \binom{v}{2}^2 \left( 1/2+e^{-r D( F_{\tmop{tol}} \| \frac{1+1/\sqrt{2}}{2} )} \right)^n .
\]
\end{theorem}

The proof of this statement goes as follows.
Since abstractly cv-qtickets consist of a set of randomly produced states and questions requested on these states the formalism of quantum retrieval games (QRGs) provides adequate modeling.
This framework is presented in a largely self-contained manner, since its generality and potential make it of independent interest.
We first provide basic definitions for QRGs and derive some simple results.
Then we present possible ways of composing QRGs together with associated quantitative bounds.
The first results are then applied to the qubit pair constituents of cv-qtickets to bound the holders potential to provide answers to complementary question.
Cv-qtickets are then modeled by a QRG for scenarios in which the holder of a cv-qticket wishes to simultaneously answer questions from two independent verifiers without any additional aid.
Finally, a combinatorial bound, similar to the one used for qtickets, is used to provide an upper limit on how the double verification probability may increase with the number $v$ of verification attempts.

\subsection{Quantum retrieval games}

Quantum retrieval games (QRGs), recently defined by Gavinsky \cite{gavinsky_quantum_2011} provide a framework to analyze protocols in which information is to be extracted from a state produced according to a classical probability distribution.
We will here present a definition of QRGs following Gavinsky as well as some additional results derived which may be of independent interest.

Alice prepares a normalized state $\rho_s=\varrho(s)/p_s$ according to the probability $p_s:=\trace[\varrho_s]$ and transfers it to Bob.
Whereas Alice remembers the index $s$ of the generated state, Bob is only provided with $\rho_s$ and a full description of the distribution from which it was generated.
Alice then asks Bob a question about $s$ which Bob attempts to answer as best as possible.
A simple possibility is for Alice to directly ask Bob the value of $s$.
In general however, the set of possible answers $A$ need not coincide with the set of indexes $S$ over the possible prepared states.
If each answer $a$ is either correct or incorrect the question may be modeled as $\sigma \in S\times A\rightarrow \{0,1\}$.
This is, $\sigma(s,a)=1$ iff the answer $a$ is correct for state index $s$ and $\sigma(s,a)=0$ otherwise.
This definition faithfully represents Gavinsky's QRGs.
We extend this notion to weighted quantum retrieval games (WQRGs) to model situations where some answers are ``more correct'' than others.
Here for each prepared state $s$ and possible answer $a$ the game will assign a non-negative real value $\sigma(s,a)$ associated to the utility function of answer $a$ given input $s$ (i.e. $\sigma \in S\times A\rightarrow \mathbbm{R}_+$).

Bob needs to choose an answer $a\in A$ and may use his copy of state $\rho_s$ to do so.
The most general strategy that Bob can take according to the laws of quantum mechanics is to perform a positive operator valued measurement (POVM).
We will consider post-selected POVMs, as opposed to a physical POVM, as those which may fail to produce a measurement outcome.
This is, whereas a physical POVM always produces an outcome from the expected set, for post-selected POVM some ``invalid'' outcomes are discarded and excluded from statistics.

In order to express the random preparation of states by Alice we first define the notion of an indexed ensemble.
\begin{definition}[Indexed ensemble]
 We will say that $\varrho$ is an ensemble on $\mathcal{H}$ indexed over $S$ iff $\forall s\in S: \varrho(s)$ is a positive operator on $\mathcal{H}$ and $\sum_{s\in S}\trace[\varrho(s)]=1$.
\end{definition}
Note that if $\varrho$ is an indexed ensemble, then $\rho=\sum_s \varrho(s)$ is a normalized density matrix.
Although Alice gives a specific state $\varrho(s)/\trace[\varrho(s)]$ to Bob, since Bob does not know $s$, he does not know which one has been received.
The state $\rho= \trace_{Alice}[ \sum_{s\in S}  s\otimes\varrho(s)]$ will be called the reduced density matrix of $\varrho$ since it corresponds to tracing out Alice's classically correlated subsystem containing the index $s$.
Without loss of generality, $\rho$ can be assumed to be full rank on $\mathcal{H}$.

In other words, a physical/selective projection $\mathcal{P}$ indexed over $A$ is simply a physical/post-selected POVM equipped with an interpretation for each possible measurement outcome in terms of possible answers in $a\in A$.
\begin{definition}[Selective and physical projections]\label{def:projections}
  We will say that $\mathcal{P}$ is a selective projection indexed over $A$ iff
  $\forall a\in A$, $\mathcal{P}(a)$ are bounded positive semidefinite operators on ${\mathcal H}$.
  It will also be a physical projection iff $\sum_a \mathcal{P}(a) = \mathbbm{1}$.
\end{definition}
Note that no normalization has been imposed for selective projections since induced probability distributions are normalized a posteriori.
An indexed ensemble and a projection on the same Hilbert space induce a joint probability distribution over the indexes $S\times A$ of prepared states and provided answers.
\begin{definition}[Induced probability distribution]
 Let $\varrho$ be an ensemble on $\mathcal{H}$ indexed over $S$ and let $\mathcal{P}$ be a projection on $\mathcal{H}$ indexed over $A$.  Then 
 \begin{equation}\label{eq:Induced probability distribution}
p(s_0,a_0)= \frac{\trace[\mathcal{P}(a_0)  \varrho(s_0)]}{\sum_{s, a} \trace[\mathcal{P}(a)  \varrho(s)]}. 
\end{equation}
is a probability distribution over $S\times A$ which will be denoted by $p=\langle \varrho, \mathcal{P} \rangle$ and is undefined unless $\sum_{s,a} \tmop{Tr} [\mathcal{P}(a)  \varrho(s)] > 0$.
\end{definition}
Furthermore, note that for physical projections the denominator in Eq. (S\ref{eq:Induced probability distribution}) is 1 and the marginal of the resulting distribution over $S$ is $p(s)=\sum_a p(s,a)=\trace[\varrho(s)]$ which is independent of $\mathcal{P}$.

\begin{definition}[Weighted quantum retrieval games]\label{def:quantumWeightedRetrievalGames}
  Let $\varrho$ be an ensemble on $\mathcal{H}$ indexed over $S$.
  Consider a utility function $\sigma \in S\times A\rightarrow \mathbbm{R}_+$.
  Then the pair $\mathcal{G}= (\varrho, \sigma)$ is a weighted quantum retrieval game.
  A WQRG is also a QRG when $\sigma \in S\times A\rightarrow \{0,1\}$. 
\end{definition}

The value of a game $\mathcal{G}$ w.r.t. a projection $\mathcal{P}$ is the average utility obtained by Bob by using a certain measurement strategy $\mathcal{P}$.
This value is given by the expectancy of the utility function $\sigma$ over the joint distribution of prepared states and measurement outcomes.
\begin{definition}\label{def:ValueOfGameRespectToProjection}
  The value of game $\mathcal{G}=(\varrho, \sigma)$ w.r.t. projection $\mathcal{P}$ is defined as
  \begin{equation}\label{eq:ValueOfGameProjection}
    \tmop{Val}(\mathcal{G},\mathcal{P}) := \sum_{s,a} p(s,a)\sigma(s, a)
  \end{equation}
   where $p=\langle \varrho, \mathcal{P} \rangle$ is the induced probability distribution.
\end{definition}
We now define the optimum value achievable by Bob for two distinct conditions depending on whether selective or physical projections are allowed.
\begin{definition}\label{def:SelectiveAndPhysicalValues} The selective (respectively physical) value of a game $\mathcal{G}$ are defined as
 \begin{equation}
  \tmop{Sel}(\mathcal{G}) := \sup_{\mathcal{P} \in \text{Selective projections}} \tmop{Val}(\mathcal{G},\mathcal{P})
 \end{equation}
 \begin{equation}
  \tmop{Phys}(\mathcal{G}) := \sup_{\mathcal{P} \in \text{Physical projections}} \tmop{Val}(\mathcal{G},\mathcal{P}).
 \end{equation}
\end{definition}
Note that according to this definition $\tmop{Sel}(\mathcal{G}) \geq \tmop{Phys}(\mathcal{G})$ since the supremum is taken over a larger set.
However, for certain tailored games, the selective and physical values will coincide.
The advantage of selective values is that they may be straightforwardly computed and are more amenable to compositional results.
If Bob is forced to provide an answer, he can only achieve the physical value of a game.
If Bob is allowed to abort the game after measuring his state $\rho_s$ and aborted games are not considered when calculating his expected utility then he will be able to achieve the selective value.

The following result provides an explicit formula to calculate the selective value of a game.
In this sense, it is a generalization of lemma 4.3 in \cite{gavinsky_quantum_2011}.
\begin{theorem}[Selective value of a game]\label{thm:SelectiveValueofGames}
  Let $\mathcal{G}=(\varrho,\sigma)$ be a WQRG with $\sum_s\varrho(s)=\rho$.
  Define $O(a) := \sum_s \sigma(s, a) \rho^{-1/2} \varrho(s) \rho^{-1/2}$. 
  Then the selective value of $\mathcal{G}$ may be calculated as $\tmop{Sel} (\mathcal{G}) = \max_a \| O(a) \|$, where $\|\cdot\|$ denotes the operator norm.
\end{theorem}
\begin{proof}
  We first use the definition of the value of a game $\mathcal{G}$
  w.r.t. $\mathcal{P}$, expand the induced probability distribution and move the sum over $s$ inside the trace
  \begin{equation}
    \tmop{Val}(\mathcal{G}, \mathcal{P}) = 
    \frac{\sum_a \tmop{Tr} [\mathcal{P}(a) \sum_s \sigma(s, a) \varrho(s) ]}{\sum_a
    \trace[ \mathcal{P}(a) \sum_s \varrho(s)]}.
  \end{equation}
  We define $\tilde{\mathcal P}$ such that $\tilde{\mathcal P}(a) = \rho^{1/2}\mathcal{P}(a)\rho^{1/2}$. 
  Using this definition and that of $\rho$ and $O_a$ we may rewrite
  \begin{equation}
  \begin{split}
    \tmop{Val} ( \mathcal{G}, \mathcal{P}) &= \frac{\sum_a \tmop{Tr}
    [\tilde{\mathcal P}(a) O(a)]}{ \sum_a \tmop{Tr} [\tilde{\mathcal P}(a)]} \\
    &\leq \max_a \frac{\tmop{Tr} [\tilde{\mathcal P}(a) O(a)]}{\tmop{Tr} [\tilde{\mathcal P}(a)]} \\
    &\leq \max_a \|O(a) \|.
  \end{split}
  \end{equation}
  The first inequality uses the positivity of all summands.
  For the second inequality we note that $\tilde{\mathcal P}(a)$ must be positive semidefinite and the variational definition of operator norm of the positive semidefinite operator $O(a)$. 
  Equality can be achieved by taking $\tilde{\mathcal P}(a_0)$ to be a projector onto the highest eigenvalue subspace of  $O(a_0)$ if $\|O(a_0) \|=\max_a \|O(a) \|$ and taking $\tilde{\mathcal P}(a_0)=0$ otherwise.
\end{proof}

The theorem provides an explicit construction of a projection achieving the selective
value of a game.
Furthermore, the proof allows us to derive a necessary and sufficient condition under which the selective and physical values of a game coincide.

\begin{corollary}
  Given a retrieval game $\mathcal{G}$, we have that $\tmop{Sel}(\mathcal{G})=\tmop{Phys}(\mathcal{G})$ iff there exist positive $\tilde{\mathcal P}(a)$ such that
  \begin{equation}
    \hspace{1em} O(a) \tilde{\mathcal P}(a) = \tmop{Sel} (\mathcal{G}) \tilde{\mathcal P}(a) \hspace{1em} \tmop{and}
    \hspace{1em} \sum_{a} \tilde{\mathcal P}(a) = \rho
  \end{equation}
\end{corollary}

We now turn to the systematic composition of retrieval games in the form of product and threshold games. 
This provides a way to construct more elaborate retrieval games together with bounds on their associated values.
A natural definition of tensor product may be given for indexed ensembles, projections and utility functions.
\begin{align}
 (\varrho_1 \otimes \varrho_2)(s_1,s_2) &=\varrho_1 (s_1) \otimes \varrho_2(s_2)\\
 ({\mathcal P}_1 \otimes {\mathcal P}_2)(a_1,a_2) &= {\mathcal P}_1 (a_1) \otimes {\mathcal P}_2(a_2)\\
 (\sigma_1\otimes\sigma_2)((s_1, s_2),(a_1, a_2)) &= \sigma_1(s_1,a_1)\sigma_2(s_2,a_2)
\end{align}
These definitions have the property that the tensor product of physical projections is a physical projection and that the induced probability distribution of two tensor product is the tensor product of the individual induced probability distributions
\[
\langle (\varrho_1 \otimes \varrho_2),({\mathcal P}_1 \otimes {\mathcal P}_2)\rangle = \langle \varrho_1 ,{\mathcal P}_1 \rangle \otimes \langle \varrho_2,{\mathcal P}_2 \rangle 
\]

\begin{definition}[Tensor product WQRG] 
Let $\mathcal{G}_{1}=(\varrho_1,\sigma_1)$ and $\mathcal{G}_2=(\varrho_2,\sigma_2)$. 
We define the tensor product WQRG $\mathcal{G}_1\otimes \mathcal{G}_2$ as
\[
\mathcal{G}_1\otimes \mathcal{G}_2 = (\varrho_1\otimes\varrho_2, \sigma_1\otimes\sigma_2).
\]
\end{definition}

\begin{proposition}[Tensor product selective value]\label{prop:SelectiveValueofTensor}
The selective value of a tensor product game is the product of the selective value of the independent games.
\[
  \tmop{Sel}(\mathcal{G}_1\otimes \mathcal{G}_2) = \tmop{Sel}(\mathcal{G}_1)\tmop{Sel}(\mathcal{G}_2)
\]
\end{proposition}
\begin{proof}
 By using the definition of $O(a)$ in theorem \ref{thm:SelectiveValueofGames} with respect to the WQRG involved we obtain
 \[
\| O(a_1, a_2)\| = \| O_1(a_1) \otimes O_2(a_2) \| = \| O_1(a_1) \| \| O_2(a_2) \|.
 \]
 Maximizing over $a_1$ and $a_2$ on both sides theorem \ref{thm:SelectiveValueofGames} provides the desired equality.
\end{proof}

The selective value of the product game is attained by the tensor product of projections, each achieving the respective selective values.

\begin{corollary}[Tensor product physical value] 
If $\tmop{Phys}({\mathcal G}_1)=\tmop{Sel}({\mathcal G}_1)$ and $\tmop{Phys}({\mathcal G}_2)=\tmop{Sel}({\mathcal G}_2)$ then $\tmop{Phys}({\mathcal G}_1\otimes {\mathcal G}_2)=\tmop{Sel}({\mathcal G}_1\otimes {\mathcal G}_2)$.
\end{corollary}


Given a direct product game and a projection for it one may consider the inverse procedure of defining a projection on one of the subcomponents of the game.
\begin{definition}[Restriction of a projection]
Let $\mathcal{P}$ be a projection on $\mathcal{H}_1\otimes\mathcal{H}_2$ indexed over $A_1\times A_2$.
Furthermore, let $\rho_2$ be a normalized density matrix on $\mathcal{H}_2$.
We define the restriction $\mathcal{P}_{|1}$ with respect to $\rho_2$ and $A_2$ as
 \[
 \mathcal{P}_{|1}(a_1) = \sum_{a_2} \trace_2 (\mathcal{P}(a_1,a_2) \mathbbm{1}\otimes\rho_2).
 \]
\end{definition}
By abuse of notation, if $\rho=\rho_1\otimes\rho_2$ is a normalized product state in $\mathcal{H}_1 \otimes \mathcal{H}_2$ we may define the restriction of $\mathcal{P}$ with respect to the normalized tensor factors of $\rho$.
This is the case for the reduced density matrix of product indexed ensembles.
By restricting a projection one obtains a new projection which induces the same reduced probability distribution

\begin{lemma}[Restriction of a projection]
Let $\mathcal{P}_{|1}$ be the restriction of $\mathcal{P}$ with respect to $\rho_2$ and $A_2$, where $\rho_2$ is the reduced density matrix of $\varrho_2$.
Then 
 \[
 \langle \varrho_1, \mathcal{P}_{|1} \rangle(s_1,a_1) = 
 \sum_{s_2,a_2}\langle \varrho_1\otimes\varrho_2, \mathcal{P} \rangle(s_1s_2,a_1a_2).
 \] 
\end{lemma}

\begin{theorem}[Selective value of threshold QRG]\label{thm:ThresholdWQRG}
Let $\mathcal{G}_j=(\varrho_j,\sigma_j)$ be WQRGs s.t. $\sigma_j \in (S_j,A_j)\rightarrow [0,1]$ and $\tmop{Sel}(\mathcal{G}_j) = \delta_j$ for all $j\in\{1,\ldots,n\}$.
Furthermore take $\delta = n^{-1}\sum_{j=1}^n \delta_j$ and $\delta\leq \gamma \leq 1$.
Define the QRG $\mathcal{G_{\gamma}}=(\bigotimes_j \varrho_j), \sigma_{\gamma} )$ with a tensor product ensemble distribution and boolean utility function 
\[
 \sigma_\gamma(\vec{s},\vec{a}) =  \left( \sum_{j=1}^n \sigma_j(s_j,a_j) \geq \gamma n \right).
\]
Then we have $\tmop{Sel}(\mathcal{G_\gamma})\leq  2 e^{-nD(\gamma \| \delta)} $.
\end{theorem}
\begin{proof}
The direct product indexed ensemble $\varrho=\bigotimes_j \varrho_j$ and projection $\mathcal{P}$ induce a normalized probability distribution over $\vec{S}\times \vec{A}$ given by 
\[
p(\vec{s},\vec{a}) = \frac{\trace[\mathcal{P}(\vec{a}) \varrho(\vec{s})]}{\sum_{\vec{s}\vec{a}}\trace[\mathcal{P}(\vec{a}) \varrho(\vec{s})]}. 
\]
Define the dependent random variable $X_j$ to be $\sigma_j(s_j,a_j)$ where $s_j$ and $a_j$ are taken according to this probability distribution.
For any $S \subseteq \{1,\ldots,n\}$, we may define $\mathcal{P}_{|S}$ as the restriction of the projection $\mathcal{P}$ to the subsystems specified by $S$ with respect to $(\rho_{\vec{s}})$.
By proposition \ref{prop:SelectiveValueofTensor} we have that 
\begin{equation}
\Exp\left[  \prod_{j\in S} X_j\right ] = \tmop{Val}\left(\bigotimes_{j\in S} \mathcal{G}_j, \mathcal{P}_{|S}\right) \leq \prod_{j\in S}\delta_j.  
\end{equation}
Using theorem \ref{thm:GeneralizedChernoffBound} and definition \ref{def:ValueOfGameRespectToProjection} we obtain
\begin{equation}
 \tmop{Val}\left( \mathcal{G}_\gamma , \mathcal{P} \right) =\Pr\left[ \sum_j X_j \geq \gamma n\right]\leq 2 e^{-nD(\gamma \| \delta)}.
\end{equation}
Since this is true for arbitrary $\mathcal{P}$ we conclude that $\tmop{Sel}(\mathcal{G}_\gamma) \leq 2 e^{-nD(\gamma \| \delta)}$.
\end{proof}

\subsection{cv-qticket qubit pair building block}

Consider a game in which Alice transfers to Bob one of the following states chosen at random
\[
S=\{ \vert 0, + \rangle, \vert 0, - \rangle, \vert 1, + \rangle, \vert 1,- \rangle,
\vert +, 0 \rangle, \vert -, 0 \rangle, \vert +, 1 \rangle, \vert -,1 \rangle \},
\]
each with probability 1/8.
Alice then asks Bob for the $Z$ polarization of both qubits, possible answers being $A=\{00,01,10,11\}$.
An answer is correct iff it coincides in the polarization of the qubit prepared in a $Z$ eigenstate.
Bob can always answer the question correctly by measuring both qubits in the $Z$ basis.

The quantum retrieval game formalism applies to this problem although one must admit that it is like cracking a nut with a sledgehammer.
We call this game $\mathcal{G}_Z = (\varrho, \sigma_Z)$ where we have $\sum_s \varrho(s) = \rho = \mathbbm{1}_4/4$, and $\trace(\varrho(s))=1/8$ for all $s\in S$.
A formal definition of the utility function $\sigma_Z$ can be given as $\sigma_Z(s,a) = \left( s_1\equiv a_1 \tmop{ or } s_2\equiv a_2 \right )$.
We first define the operators $O(a)$ from theorem \ref{thm:SelectiveValueofGames}.
Due to symmetry we may restrict to considering one such operator
\begin{equation}
 O(00)= 4 \left(  \varrho(0,+) + \varrho(0,-) + \varrho(+,0) +\varrho(-,0) \right)
\end{equation}
and find that $\|O(00)\|=1$ which is a non degenerate eigenvalue for all $O(a)$.
The fact that the four corresponding eigenspaces are orthogonal confirms that 1 is also the physical value of the game.

The same trivial value of 1 can be achieved for the game in which Alice requests the $X$ direction polarization of the states.
We will call this game $\mathcal{G}_X = (\varrho, \sigma_X)$.
The problem becomes interesting if Bob is requested provide a guess for both complementary polarizations.
There are two relevant possibilities, both of which will require Bob to give an answer twice as long as before.
The first scenario describes the best case probability of Bob answering both questions correctly and may be modeled by a QRG with utility function
\[
\mathcal{G}_\wedge=(\varrho,\sigma_\wedge) \quad\sigma_\wedge(s,a_X a_Z) = \sigma_X(s,a_X) \wedge \sigma_Z(s,a_Z).
\]
In the second scenario we are interested in the average number of questions answered correctly when two complementary questions are posed and may be modeled by the WQRG with utility function
\[
\mathcal{G}_{\tmop{avg}}=(\varrho,\sigma_{\tmop{avg}})\quad
\sigma_{\tmop{avg}}(s,a_X a_Z) = \frac{\sigma_X(s,a_X) + \sigma_Z(s,a_Z)}{2}.
\]
Thanks to symmetries one need only calculate a single $\| O(a) \|$  and for concreteness we choose $O(++00)$.
For the conjunction QRG we obtain
\[
O(++00) = 4 \left(  \varrho(0,+) + \varrho(+,0) \right)\text{ and } \|O_{++00}\|=3/4.
\]
For the average WQRG we obtain
\begin{equation}
\begin{split}
O(++00) = & 2 [  2\varrho(0,+)  + 2\varrho(+,0) + \varrho(0,-)\\
  & + \varrho(-,0) + \varrho(+,1) + \varrho(1,+) ]  
 \end{split}
\end{equation}
and $\|O_{++00}\|=1/2+1/\sqrt{8} \approx 0.8536$.
This is precisely the optimal fidelity for covariant qubit cloning (i.e. cloning of equatorial qubits).
On the other hand, if Bob is asked the same question twice instead of complementary questions it is clear that he will be able to repeat two correct answers.
All in all, if Bob is asked complementary question half of the time and coinciding questions half of the time he will be able to emulate an average fidelity of $3/4+\sqrt{2}/8\approx 0.927$.

Indeed, once we have defined a concrete WQRG, calculating its selective value becomes an exercise thanks to theorem \ref{thm:SelectiveValueofGames}.
Furthermore, if the game has sufficient symmetry it will be possible to prove a coinciding physical values for the game.

\subsection{cv-qticket}

We will first bound the probability of answering two of these randomly chosen questions by bounding the selective value of the corresponding retrieval game.
To do this, we bound the value of a game where $r$ complementary questions are asked on $r$ qubit pairs (this is precisely the case for one block when the two random questions are complementary).
\begin{equation}
\begin{split}
\sigma^{(X)}_{F_{\tmop{tol}}}(\vec{s},\vec{a}^{(X)}) =& \left( \sum_{j=1}^r \sigma^{(X)}_j(s_j,a^{(X)}_j) \geq F_{\tmop{tol}} r \right) \\
\sigma^{(Z)}_{F_{\tmop{tol}}}(\vec{s},\vec{a}^{(Z)}) =& \left( \sum_{j=1}^r \sigma^{(Z)}_j(s_j,a^{(Z)}_j) \geq F_{\tmop{tol}} r \right) \\
\sigma^{\wedge}_{F_{\tmop{tol}}} (\vec{s}, (\vec{a}^{(X)},\vec{a}^{(Z)})) =& \sigma^{(X)}_{F_{\tmop{tol}}} (\vec{s}, \vec{a}^{(X)}) \wedge \sigma^{(Z)}_{F_{\tmop{tol}}} (\vec{s}, \vec{a}^{(Z)})
\end{split}
\end{equation}

We will not calculate the selective value exactly but give a bound in terms of theorem \ref{thm:ThresholdWQRG}.
In order for the two block answers to be correct, among the two, at least $2F_{\tmop{tol}}r$ answers should have been provided correctly for individual qubit pairs.
This is a weaker condition since it only imposes that the sum among the two block answers be sufficiently large, not necessarily implying that they are both above threshold.
\begin{equation}
 \sigma^{\wedge}_{F_{\tmop{tol}}} (\vec{s}, (\vec{a}^{(X)},\vec{a}^{(Z)})) \leq \left( \sum_{j=1}^r \sigma^{avg}_j(s_j,(a^{(X)}_j,a^{(Z)}_j)) \geq F_{\tmop{tol}} r \right)
\end{equation}
The description on the right hand side has precisely the form required for theorem \ref{thm:ThresholdWQRG}.
We conclude that the selective value and hence the probability within any strategy of providing valid answers to two complementary questions for the same block is upper bounded by $2\exp[-rD(F_{\tmop{tol}} \| 1/2+1/\sqrt{8})]$ (for $F_{\tmop{tol}}>1/2+1/\sqrt{8}$).

Given two randomly chosen questions for a block there is a probability of $1/2$ that they will coincide and a probability $1/2$ that they will be complementary.
Taking this into account, the probability for a dishonest holder to correctly answer two such randomly chosen block questions is upper bounded by $1/2+\exp[-rD(F_{\tmop{tol}} \| 1/2+1/\sqrt{8})]$.
By taking $r$ sufficiently large, this value can be guaranteed to be smaller then 1.
Hence, the probability of correctly answering $n$ such randomly chosen threshold question pairs will be upper bounded by $B:=(1/2+\exp[-rD(F_{\tmop{tol}}, 1/2+1/\sqrt{8})])^n$ which can be made exponentially close to 1 in $n$.

\subsection{Combinatorial bound on choosing and learning}

The formulation presented adequately models a scenario in which the holder of a cv-qticket does not receive any feedback from the verifiers.
However, if the holder of a cv-qticket can engage in several verification protocols, new possibilities arise which should be taken into account.

Firstly, by simultaneously engaging in several ($v$) verification protocols with different verifiers, the holder may simultaneously have access to $v$ challenge questions.
The holder may then for instance, choose the most similar questions and attempt to answer these.
Furthermore, by successively participating in $v$ verification protocols the holder can choose to perform verifications sequentially and wait for the outcome of the $k$-th before choosing which question to answer as the $k+1$-th and providing an answer for it.

In general, if the holder engages in $v$ verification attempts, he will receive $v$ random questions providing no additional information on the cv-qticket. 
There are $\binom{v}{2}$ possible question pairs among these, each of which can be seen as randomly chosen. 
Thus if no feedback is used the probability of answering at least one of these pairs correctly is upper bounded by $\binom{v}{2}B$.
An example scenario where this bound is relatively tight is when $r$ is very large and $n$ is relatively small. 
In this case, the probability of answering two randomly chosen questions is well approximated by the collision probability $2^{-n}$ ( i.e. the probability that two questions coincide ) which grows precisely as $\binom{v}{2}$ if the holder has access to $v$ independently drawn questions and may choose to answer any pair.

Suppose now, that the answers to the verifiers are provided sequentially, so that the decision of which answer to produce for each verifier may be made dependent on the outcome of previous verifications.
We can safely assume that the answers to challenge questions are then provided sequentially, each after receiving the acceptance or rejection of the previous ones.
We can then apply a similar argument to the one exposed for the proof of qticket security in section \ref{sec:CombinatorialBoundOnLearning}.
This yields an additional factor of $\binom{v}{2}$ corresponding to the possible feedback scenarios up to the point of the second accepted answer, each of which can be simulated statically (i.e. by assuming the given feedback and fixing a corresponding POVM to generate answer up to that point).
Hence the total probability for an interactive strategy with $v$ verification attempts of producing two or more accepted answers is upper bounded by $\binom{v}{2}^2 B$.

It may seem artificial for verifiers to select a random question each time.
Randomness is important in order to avoid revealing information about the issued cv-qticket.
However, the verifier may choose a random question once and for all and ask it until it is answered correctly.
Once it has been answered correctly, the verifier knows that the cv-qticket has already been redeemed and can thus reject all subsequent verification attempts.
This is similar to the kind of scheme used for prepaid telephone cards discussed in the applications section.
However, the quantum case provides an advantage since one may have multiple verifiers which do not communicate.
In a simple example with two verifiers, two composite questions may be chosen such that they are complementary on every qubit pair (i.e. one question is chosen at random and uniquely determines the other).

\section{Applications}

Our quantum information application attempts to reduce quantum requirements to a minimum.
However, even prepare and measure qubit memories remain technologically challenging.
For problems admitting a classical solution, such an approach is likely to be technologicaly less demanding.
In other words, relevant applications for prepare and measure quantum memories will be those solving problems for which no classical solutions are known.
In this section we discuss some problems with classical solutions and propose refinement of such problems for which no classical solution is possible.

\subsection{ Enforcing single usage with a single verifier }
For some applications, the no cloning of quantum information is only an apparent advantage.
Our qticket and cv-qticket constructions can guarantee an exponentially small double usage probability.
However, this is not an impresive feat for scenarios where there is a single verifier or if the verifiers have acces to realtime communication with a centralized database.
In this case, a randomly chosen classical ticket has equaly good properties.
After a ticket is succesfully redeemed once, it can be removed from the central database, making it invalid for any succesive verification attempt.
In fact this classical strategy is widely used for crediting prepaid phone lines with a client calling a toll free number and typing the purchased ticket number in order to credit a telephone account.
Thus in such scenarios, the quantum strategy does not provide additional protection with respect to a classical solution.

\subsection{Multiple non communicating verifiers}

In scenarios with multiple non communicating verifiers, (cv-)qtickets provide a soution to a problem where all classical approaches fail.
We describe a {\it witness protection program} as an example of how such a scenario might look like.

In a witness protection program, a governmental institution decides to give asylum to a key eye witness to whom an unforgeable quantum token is issued.
This token can be used by the witness (holder) to claim asylum in any of a set of participating hotels (verifiers).
The issuer also provides all hotels with the necessary information to verify the tokens.
When using the token, neither the eye-witness nor the chosen hotel wish to divulge the locale where the witness is hosted, thus protecting both from being targets of an attack.
This includes suspending communication between participating hotels as well as with the issuing authority.
Any classical solution can not prevent a sufficiently resourceful holder from making copies of the received token, thus hotels are forced to communicate in order to avoid its double use. 
In this case, a quantum solution based on unforgeable tokens is the sole possibility to satisfy these unique constraints.
\begin{figure}[ptb]
\includegraphics[width=0.9\columnwidth]{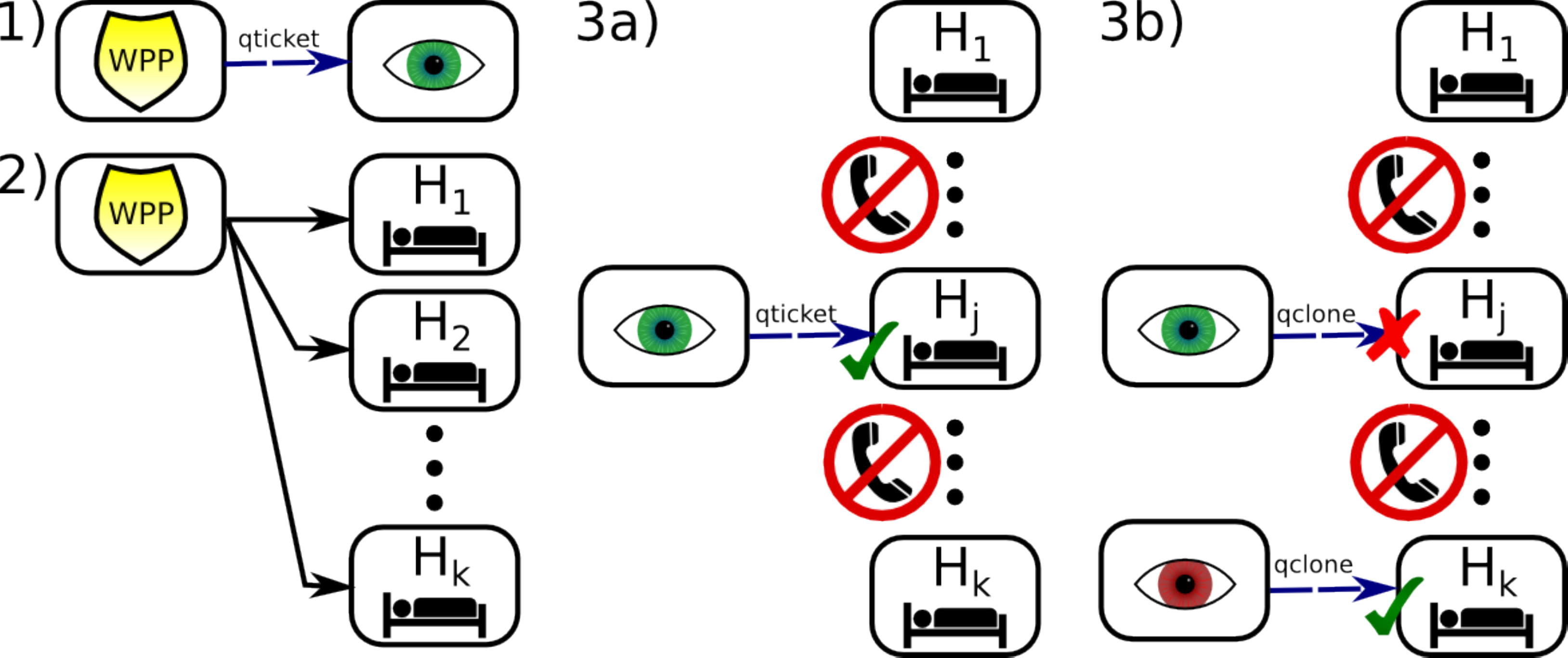}
\caption{
1) The issuing entity hands a qticket to the key witness. 2) It provides the hotels with the secret classical description which will be used to verify it. 3a) An honest witness choses a hotel and physically transfer the qticket for verification. It will be accepted as long as the level of accumulated noise is below threshold. 3b) A dishonest witness will fail to counterfeit his/her qticker to provide acomodation for an additional guest. However, there is no way of avoiding a valid qticket from changing hands.}
\label{fig:WitnessProtection}
\end{figure}

\subsection{Reduced availability under sporadic verification}

In principle, a centralized database may guarante that classical ticket are only redeemed once. 
However, there are situations where the ticket should be available only to one holder at a time and the non-clonable nature of a qticket allows enforcing this.
One such example is the sporadic control of tickets required for a given service.
For concreteness, imagine a qticket which is valid for making use of a public transportation network.
Commuters are sporadically controled, at which point if they are found to have a invalid qticket they are charged an important fine, whereas if they are found to hold a valid qticket, they are provided with a fresh substitute.
If the transportation tickets are classical, sporadic control can not avoid counterfeited copies in the hands of coluding commuters from circulating simultaneously.
The deceiving commuters need only communicate classicaly among each other before and after they are controled, effectively sharing a single classical ticket to make use of the service multiple times\footnote{If the classical ticket is not renewed upon control even communication is unnecesary.}.
In contrast the unavailability of long distance quantum comunication would disallows them to share a qticket in such a way (i.e. each valid qticket may only be at one place at a time).

\subsection{The Quantum Credit Card}

Having developed a single verification, noise tolerant, non-forgeable token, such as the cv-qticket, it is now possible to examine generalizations to interesting composite protocols.
For instance, Gavinsky's proposal\cite{gavinsky_quantum_2011} allows for multiple verification rounds to be performed on a single token, while also ensuring that the token can not be split into two independently valid subparts.
Such a construction may be seen as a quantum credit card. Indeed, the classical communication which takes place with the issuer (bank) to verify the cv-qticket (via ``challenge'' questions) may be intentionally publicized to a merchant who needs to be convinced of the card's validity.
An alternate possibility is to follow the original interpretation as a quantum cash token where verification is performed by the receiver each time the ``money'' changes hands.

\subsection{ Excluding eavesdropers }
While qtickets do not provide additional advantage against dishonest holder in the scenario of a single verifier quantumness may provide an advantage against eavesdroping and untrusted communication.
In order to make online banking more secure, Banks routinely use TANs (transaction authentication numbers) as an additional security measure.
The bank sends its client a list of TANs via postal service in addition to an online password which is set up via another channel.
Each time a bank transaction is requested online by the client, the bank requests a TAN from the list to guarantee the authenticity of the transaction.
An impostor then needs to know both a secret password used by the user and some TANs, thus increasing the difficulty to succesfully impersonate a transaction with respect to any single security measure.
However, since TANs are classical objects it is conceivable that an eavesdroper may learn them while remaining undetected (imagine an eavesdroper taking xray pictures of the correspondence).
This means that with some effort of the eavesdroper the additional security measure becomes ineffective.

This problem can be straightforwardly resolved by using quantum prepare and measure memories.
Even if a cv-qticket is sent via an untrusted optical fiber or postal service, the receiver may openly communicate with the issuer and sacrifice some of the received qubits in order to obtain a bound on how much information could have leaked to eavesdropers.
This is precisely the approach taken in QKD to obtain a statistical bound on the information that has leaked out.
Gavinsky's $\mathcal{Q}$ scheme, allowing multiple verification rounds may be reinterpreted as quantum TAN lists.
The holder of a quantum TAN list may verify its validity, and perform a transaction by publicly communicating with the bank.
If the quantum TAN list is verified to be legitimate, then the probability of an eavesdroper getting verified by using the leaked information will be negligible (exponentially small).
In turn, the cv-qtickets described in the main text and appendix may be used as basic building blocks  for such a scheme in the precense of noise.

\subsection{Combining with classical computational assumptions}

The qticket and cv-qticket protocols proposed require the verifiers to have acces to secret information (i.e. information which is sufficient to produce additional tokens).
This poses a problem for situations where the verifiers can not be trusted with such sensitive information.
Idealy, it should be possible to verify the tokens relying exclusively on publicly available information.
Seeking to achieve this, Bennet et. al. \cite{Bennett1983} made an ingenious proposal based on computational assumptions.
We consider a family of such proposals and show that they are not secure.

The proposal of Bennet et. al can be abstractly presented as follows.
The mint secretly keeps the solution to a publicly known instance of a computationally hard problem.
The problem instance may have been generated by the mint from the solution in the first place.
It is then possible for the mint to interactively convince anyone that it has the solution to the problem without giving away any additional information about the solution.
Such a method for convincing is known as a statistical zero knowledge proof (SZKP).
For many problems, SZKP can be set up in a {\it cut and choose} fashion.
This is, the mint ``cuts'' the secret solution into two parts $m_1$ and $m_2$ which toghether allow reconstructing the secret solution but independently provide no usefull information.

We provide a simple example for this in the context of graph isomorphism.
The mint generates two isomorphic graphs $G_1$ and $G_2$, publishes them but keeps the isomorphism secret.
The mint can then proceed as follows to convince someone that it has a proof of $G_1 \equiv G_2$. 
The mint generates a random graph $G_c\equiv G_1$ (which may be seen as the ``cut'') by randomply permuting the vertices in $G_1$ and offers to provide either the isomorphism proving $G_c\equiv G_1$ or  the one for $G_c\equiv G_2$.
A verifier then randomly ``chooses'' which of the two to request, a dishonest proover (lacking a proof of $G_1 \equiv G_2$) would have no more than a $50\%$  chance of being capable to answer correctly.
Thus by repeating this procedure $n$ times, a verifier can be $1-2^{-n}$ certain of the proof being conveyed.
On the other hand, the verifier learns no information about the secret solution $G_1\equiv G_2$ which it could not have easily obtained independently.

In our language, Bennet et al. go one step further and propose that the mint hardcode such SZKPs into a token composed of a classical part and a quantum state.
The holder of such a token would thus have access to a one time SZKP of the problem openly published by the mint.
In order to do this, the ``cut'' is provided classicaly, whereas the two possible proofs are encoded in a quantum state such that either of them may be extracted by measuring the state but not both.
As we will show however, this requirement can not be fullfiled if one allows arbitrary coherent operations to be performed on the quantum tokens (i.e. security against the most general attacks is impossible).

Suppose that we wish to encode two possible messages $m_1$ and $m_2$ into a
large quantum state such that an adversary may learn either of them with a
very high certainty but may not learn both of them.
In order to show that this is not possible, we consider the simple scenario where two bits of information are encoded in a quantum state such that either of them can be retrieved by measurement with high presicion and show that it is possible (at least in theory) to retrieve both of them simultaneously with high succes rate.

Consider a family of quantum states $\rho_{\alpha, \beta}$ used to multiplex the classical bits $\alpha,\beta \in \{0,1\}$. 
That the holder of the quantum state can extract either $\alpha$ or $\beta$ may be  expressed by
\begin{equation}
\begin{split}
   \tmop{Tr}[ P^a_{\alpha} \rho_{\alpha', \beta'} ]  \geq  (1-\epsilon)\delta_{\alpha, \alpha'} \\
  \tmop{Tr}[ P^b_{\beta} \rho_{\alpha', \beta'}]  \geq  (1-\epsilon)\delta_{\beta, \beta'}
  \end{split}
\end{equation}
where $\{P^a_{\alpha}\}$ and $\{P^b_{\beta}\}$ are two projective measurements indexed over $\alpha$ and
$\beta$ respectively (i.e. $P^a_0 + P^a_1 = P^b_0 + P^b_1 = \mathbbm{1}$). 
We wish to show that by succesively performing both projective measurements both $\alpha$ and $\beta$ may be obtained with relatively high certainty. 
Hence it is impossible to multiplex the $\alpha$ and $\beta$ such that either may be recovered reliably but attempting to measure both remains highly unreliable. 
The POVM corresponding to sequential measurement is given by
$\tilde{P}_{\alpha, \beta} = P^a_{\alpha} P^b_{\beta} P^a_{\alpha}$ and we
will show that it is good to order $\sqrt{\epsilon}$
\begin{equation}
\begin{split}
   &  \tmop{Tr}[ P^a_\alpha P^b_\beta P^a_\alpha \rho_{\alpha,\beta} ] \\
 = & \tmop{Tr}[ P^b_\beta \rho_{\alpha,\beta} ]  - \tmop{Tr}[ P^b_\beta P^a_{\bar{\alpha}} \rho_{\alpha,\beta} P^a_{\bar{\alpha}}]  - (\tmop{Tr}[P^b_\beta P^a_{\bar{\alpha}} \rho_{\alpha,\beta} P^a_\alpha ]  + c.c) \\
   \geq & 1 - 2\epsilon - 2 \sqrt{\tmop{Tr}[ P^b_\beta P^a_{\alpha} \rho_{\alpha,\beta} P^a_\alpha ]  \tmop{Tr}[ \rho_{\alpha,\beta} P^a_{\bar\alpha} ] }\\
  \geq & 1 - 2 \epsilon - 2 \sqrt{\epsilon}
\end{split}
\end{equation}
where we have used the Cauchy-Schwartz inequality to go obtain the second line
simplifying further with $P^a_{\alpha} \succ (P^a_{\alpha})^2$ and $P^b_{\beta} \succ (P^b_{\beta})^2$.
Hence if we make $\epsilon$ arbitrarily small, this will also make the probability of extracting both indices arbitrarily close to 1.